\documentclass[aps,showpacs,prd,twocolumn,nofootinbib,nobibnotes,]{revtex4-2}

\usepackage{graphicx}
\usepackage{amssymb}
\usepackage{amsmath}
\usepackage{color}
\usepackage{float}
\usepackage{ulem}
\usepackage{accents}
\usepackage{graphicx}
\usepackage{graphicx}
\usepackage{amsfonts}
\usepackage[colorlinks=true,
pdfstartview=FitV,linkcolor=blue,
citecolor=blue,urlcolor=blue,breaklinks=true]
{hyperref}
\usepackage{array}
\usepackage{float}
\usepackage{placeins}
\usepackage[dvipsnames]{xcolor}
\usepackage{csquotes}
\usepackage{bbold}
\usepackage{units}
\usepackage{enumitem}
\usepackage{dsfont}
\usepackage{upgreek}

\newcolumntype{C}[1]{>{\centering\arraybackslash}m{#1}}

\renewcommand{\eqref}[1]{\mbox{Eq.~(\ref{#1})}}

\definecolor{ForestGreen}{rgb}{0.13,0.55,0.13}

\usepackage{fixmath}

\newcommand{\orcid}[1]{\href{https://orcid.org/#1}{\includegraphics[width=10pt]{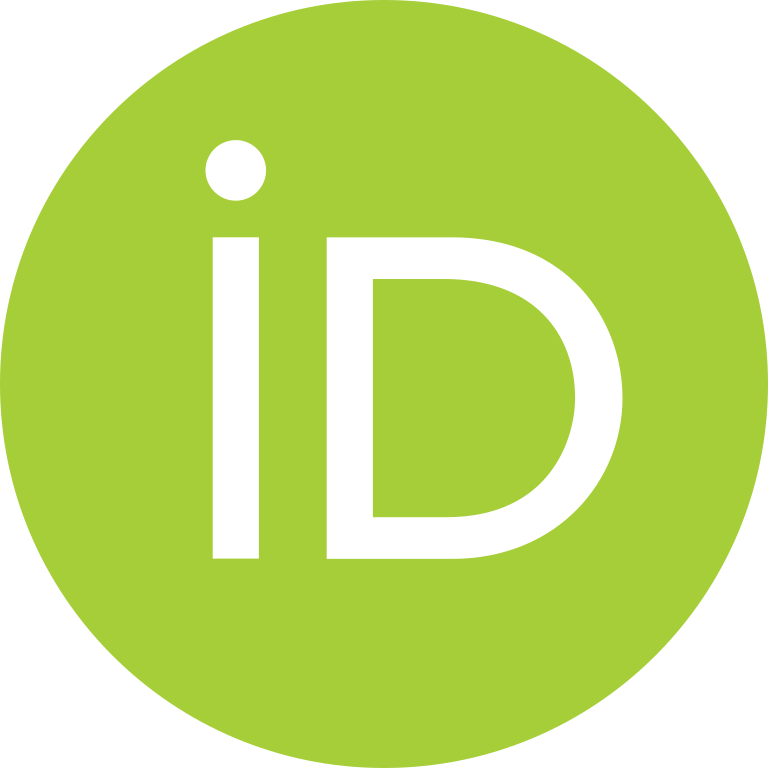}}}

\begin{document}
	
	\title{Cold plasma {modes} in the chiral Maxwell-Carroll-Field-Jackiw electrodynamics}

\author{Filipe S. Ribeiro\orcid{0000-0003-4142-4304}$^a$}
\email{filipe.ribeiro@discente.ufma.br, filipe99ribeiro@hotmail.com}
	\author{Pedro D. S. Silva\orcid{0000-0001-6215-8186}$^a$}
	\email{pedro.dss@discente.ufma.br, pdiego.10@hotmail.com}
	\author{Manoel M. Ferreira Jr.\orcid{0000-0002-4691-8090}$^b$}
	\email{manojr.ufma@gmail.com, manoel.messias@ufma.br}
		\affiliation{$^a$Programa de P\'{o}s-graduaç\~{a}o em F\'{i}sica, Universidade Federal do Maranh\~{a}o, Campus
		Universit\'{a}rio do Bacanga, S\~{a}o Lu\'is, {Maranhão} 65080-805, Brazil}
	\affiliation{$^b$Departamento de F\'{i}sica, Universidade Federal do Maranh\~{a}o, Campus
		Universit\'{a}rio do Bacanga, S\~{a}o Lu\'is, {Maranhão} 65080-805, Brazil}

	\begin{abstract}
	In this work, we study the propagation and absorption of plasma waves in the chiral
Maxwell-Carroll-Field-Jackiw  {(MCJF) electrodynamics}. The Maxwell equations are rewritten for a cold, uniform, and collisionless fluid plasma model,  allowing us to determine the new refractive indices and propagating modes. { The cases of propagation parallel and orthogonal to the magnetic field are examined} considering a purely timelike CFJ background that plays the role of the magnetic conductivity chiral parameter. {The collective electromagnetic modes are associated with} four distinct refractive indices associated with {right-circularly polarized and left-circularly
	polarized} waves. For each index, the propagation and absorption zones are illustrated for some specific parameter values. {In low-frequency regime, we have obtained modified helicons with right- and left-circularly polarizations}. The optical behavior is investigated by means of the rotatory power (RP) and dichroism coefficient. The existence of a negative refraction zone enhances the rotatory power. It is also observed RP sign reversal, a feature of rotating plasmas.	

\end{abstract}

	\maketitle

	\section{Introduction}
	
	The study of electromagnetic (EM) waves propagation \cite{refZANGWILL,refJACKSON} in cold magnetized plasma is based on magneto-ionic theory \cite{Gurnett,STURROK,Boyd, Stix, Bittencourt,Piel, chapter-8}, developed by E. Appleton \cite{Appleton32} and D. Hartree \cite{refHartree} between 1929 and 1932 to describe the radio waves propagation in the ionosphere, in the context of the usual electrodynamics \cite{RATCLIFF2}. EM waves in plasmas have been studied in other scenarios recently, as in logarithmic nonlinear electrodynamics \cite{Helayel}.	
	
The chiral magnetic effect (CME) is the macroscopic generation of an electric current in the presence of a magnetic field, stemming from an asymmetry between the number density of left- and right-handed chiral fermions \cite{Kharzeev1, Kharzeev1B, Fukushima, LiKharzeev, Vilenkin}. It has been extensively investigated in several distinct contexts, such as quark-gluon plasmas \cite{Inghirami, Schober, Akamatsu}, cosmology~\cite{Maxim}, neutron stars \cite{Leite, Dvornikov}, and electroweak interactions \cite{Maxim1}. The CME plays a very relevant role in Weyl semimetals, where it is usually connected to the chiral anomaly associated with Weyl nodal points \cite{Burkov}, the absence of the Weyl nodes \cite{Chang},  anisotropic effects stemming from tilted Weyl cones \cite{Wurff}, the CME and anomalous
		transport in Weyl semimetals \cite{Landsteiner}, quantum oscillations
		arising from the CME \cite{Kaushik}, computation of the electromagnetic
		fields produced by an electric charge near a topological Weyl semimetal with
		two Weyl nodes \cite{Ruiz}, renormalization evaluations for Weyl semimetals and Dirac materials \cite{Throckmorton}, and solutions of axion electrodynamics \cite{KDeng}.

The CME current can be classically described by the axion Lagrangian \cite{KDeng, Barnes,Wilczek, Sekine, Tobar, Paixao, Qiu},
		\begin{equation}
		\mathcal{L}=-\frac{1}{4}F^{\mu\nu}F_{\mu\nu}+\theta (\mathbf{E}\cdot \mathbf{B)},
		\end{equation}%
		where $\theta$ is the axion field. In this context, the Maxwell equations are
		\begin{align}
		\mathbf{\nabla }\cdot \mathbf{E}&=\rho -\mathbf{\nabla }\theta \cdot \mathbf{B%
		}, \label{M1A} \\
		\mathbf{\nabla }\times \mathbf{B}-\partial _{t}\mathbf{E}&=%
		\mathbf{j}+(\partial _{t}\theta )\mathbf{B}+\mathbf{\nabla }\theta \times 
		\mathbf{E},  \label{M1B}
		\end{align}
		where the terms involving $\theta$ derivatives find association with condensed matter effects \cite{Qiu}. Indeed,  $\mathbf{\nabla }\theta \cdot \mathbf{B}$ represents an anomalous charge density, while $\mathbf{\nabla }\theta \times \mathbf{B}$ appears in the anomalous Hall effect, and  $(\partial _{t}\theta )\mathbf{B}$ plays the role of the chiral magnetic current. {When we address a cold axion dark matter, the associated de Broglie wavelength is large enough to assure the inexistence of variation of the axion field in the typical dimension of experimental devices. In this case, the axion field is supposed to not depend on the space coordinates, $\mathbf{\nabla }\theta
			={\bf{0}}$, so that} the Maxwell equations (\ref{M1A}) and (\ref{M1B}) read
		\begin{equation}
		\mathbf{\nabla }\cdot \mathbf{E}=\rho, \quad
		\mathbf{\nabla }\times 
		\mathbf{B}-\partial _{t}\mathbf{E}=\mathbf{j}+(\partial _{t}\theta )\mathbf{B},
		\label{Maxwellaxion1}
		\end{equation}
		where $(\partial _{t}\theta )\mathbf{B}$, the chiral magnetic current, may also be addressed as a term of Maxwell-Carroll-Field-Jackiw (MCFJ) theory. A classical electrodynamics scenario endowed with a chiral magnetic current has been investigated considering symmetric and antisymmetric conductivity \cite{Pedro1}. The latter case has also been addressed in Ref.~\cite{Kaushik1}.
	
	The MCFJ model \cite{CFJ} is the \textit{CPT}-odd part of the \textit{U}(1) gauge sector of the Standard Model Extension (SME) \cite{Colladay}. It is described by the Lagrangian density
		\begin{align}
		\mathrm{{\mathcal{L}}}=-\frac{1}{4}F^{\mu\nu}F_{\mu\nu}-\frac{1}{4}%
		\epsilon^{\mu\nu\alpha\beta}\left( k_{AF}\right)  _{\mu}A_{\nu}F_{\alpha
			\beta}-A_{\mu}J^{\mu},\label{MCFJMATTER}
		\end{align}
		with $\left(k_{AF}\right)_{\mu}$ being the 4-vector background which controls the Lorentz violation. This theory has been investigated in multiple respects \cite{CFJ2}, encompassing radiative evaluations \cite{CFJ3, CFJ4}, topological defects solutions \cite{CFJ5}, supersymmetric generalizations \cite{CFJ6}, classical solutions, quantum aspects and unitarity analysis \cite{CFJ7}. It may also be connected with the CME in the sense that it provides a modified Amp\`ere's law,
		\begin{equation}
		\nabla\times\mathbf{B}  -\frac{\partial\mathbf{E}}{\partial t} =\mathbf{J}+  k_{AF}^{0}\mathbf{B}
		+\mathbf{k}_{AF} \times\mathbf{E} ,\label{2}
		\end{equation}
		containing the magnetic current, $\mathbf{J}_{B}=k_{AF}^{0}\mathbf{B}$, with the component  $k_{AF}^{0}$ playing the role of the magnetic conductivity. 
	
The SME photon sector is also composed of a \textit{CPT}-even term constituted of a rank 4 Lorentz-violating tensor \cite{KM}, whose components may be properly parametrized in terms of dimensionless $3\times3$ matrices, $\kappa _{DE}$, $\kappa _{DB}$, $\kappa _{HE}$, and $\kappa _{HB}$, which allow to write generalized constitutive relations between the fields $(\mathbf{D},\mathbf{E})$ and $(\mathbf{H,B})$,
		\begin{equation}
		\begin{pmatrix}
		\mathbf{D} \\
		\mathbf{H}%
		\end{pmatrix}%
		=%
		\begin{pmatrix}
		\epsilon \mathbb{1}+\kappa _{DE} & \kappa _{DB} \\
		&  \\
		\kappa _{HE} & {\mu }^{-1}\mathbb{1}+\kappa _{HB}%
		\end{pmatrix}%
		\begin{pmatrix}
		\mathbf{E} \\
		\mathbf{B}%
		\end{pmatrix}%
		\,,  \label{eq15}
		\end{equation}
		similar to the ones that hold in continuous medium electrodynamics, see Eqs.~(\ref{constitutive2a}) and (\ref{constitutive2b}). Here, $\mathbf{D}$ is the electric displacement, while $\mathbf{H}$ is the magnetic field. This \textit{CPT}-even electrodynamics was investigated in several contexts, involving consistency aspects \cite{Schreck}, finite temperature and boundary effects \cite{CPTP}. Lorentz-violating electrodynamics in continuous matter \cite{Bailey, Gomez} has been a topic of interest in the latest years due to its potential to describe interesting effects of the phenomenology of new materials, such as Weyl semimetals \cite{Marco}. A classical field theory description of wave propagation, refractive indices, and optical effects in a continuous medium described by the MCFJ electrodynamics (with usual constitutive relations), including its Lorentz-violating higher-order derivative version \cite{Leticia1}, was discussed in Ref.~\cite{Pedroo}.

Chiral media are endowed with parity violation \cite{Barron2,Hecht,Wagniere,TangPRL}, being described by parity-odd models, as bi-isotropic \cite{Sihvola} and bi-anisotropic electrodynamics \cite{Kong,Bianiso,Aladadi,Mahmood,Lorenci,Pedro3}, whose constitutive relations read
\begin{subequations}
	\label{constitutive2}
	\begin{align}
		\mathbf{D}& =\hat{\epsilon}\, \mathbf{E}+\hat{\alpha}\,
		\mathbf{B},  \label{constitutive2a} \\
		\mathbf{H}& =\hat{\beta}\, \mathbf{E}+\hat{\zeta}\,\mathbf{B },
		\label{constitutive2b}
	\end{align}
\end{subequations}
and $\hat{\epsilon}=[\epsilon_{ij}]$, $\hat{\alpha}%
=[\alpha_{ij}]$, $\hat{\beta}=[\beta_{ij}]$, and $\hat{\zeta}=[\zeta_{ij}]$
represent, in principle, $3\times 3$ complex matrices. The bi-isotropic relations involve the diagonal isotropic tensors, ${\epsilon_{ij}}=\epsilon\delta_{ij}$, ${\alpha_{ij}}
=\alpha\delta_{ij}$, ${\beta_{ij}}=\beta\delta_{ij}$. In chiral scenarios, {left-circularly polarized (LCP) and right-circularly polarized (RCP)} waves travel at distinct phase velocities, implying birefringence and optical rotation \cite{Fowles}. This phenomenon stems from the natural optical activity of the medium or can be induced by the action of external fields (e.~g., Faraday effect \cite{Bennett, Porter, Shibata}), and it is measured in terms of the rotation angle per unit length or rotatory power (RP) \cite{Condon}. Magneto-optical effects are used to investigate features of new materials, such as topological insulators \cite{Chang1, Urrutia, Lakhtakia, Winder, Li, Li1, Tse} and graphene compounds \cite{Crasee}.
	
The RP is a probe to examine the optical behavior of several distinct systems, for instance, crystals \cite{Dimitriu, Birefringence1},  organic compounds \cite{Barron2, Xing-Liu}, graphene phenomena at terahertz band \cite{Poumirol}, and gas of fast-spinning molecules \cite{Tutunnikov}. The optical rotation may depend on the frequency (RP dispersion) and undergo reversion (anomalous RP dispersion) \cite{Newnham, Tschugaeff, Tischler}. It also finds interesting applications in chiral metamaterials \cite{Woo, Zhang, Mun}, chiral semimetals \cite{Pesin, Dey-Nandy}, in the determination of the rotation direction of pulsars \cite{Gueroult2}, and in rotating plasmas, which constitutes a scenario where RP sign reversal also takes place \cite{Gueroult}. Recently, RP reversal was also reported in a bi-isotropic dielectric in the presence of chiral magnetic current  \cite{PedroPRB}. Furthermore, in the presence of absorption, dichroism is another useful tool for the optical characterization of matter. It occurs when LCP and RCP light waves are absorbed by the medium at different degrees. It has been used to distinguish between Dirac and Weyl semimetals \cite{Hosur}, perform enantiomeric discrimination \cite{Nieto-Vesperinas,Tang}, and for developing graphene-based devices at terahertz frequencies \cite{Amin}.

Another feature of chiral systems is the possible occurrence of negative refraction and negative refractive index, which was first proposed by Veselago in 1968 \cite{Veselago} and experimentally observed in 2000 \cite{Shelby,Smith-Padilla}. Later, other experiments confirmed the negative refraction by using Snell's law \cite{Parazzoli, Houck}. This unusual property was achieved in constructed metamaterials with both {negative-electric} permittivity and magnetic permeability \cite{Kadic, Engheta}. The negative refractive index also appears in quark-gluon plasmas \cite{Liu-Luo, Jamal}, magnetoelectric materials \cite{Cheng-Wei-Qiu}, metasurfaces \cite{Lei-Zhang},  chiral bi-anisotropic metamaterials \cite{Shuang-Zhang,Zhou1}, and new materials, such as Dirac semimetals \cite{Chen-Hsu, Ball}. In chiral plasmas described by generalized bi-isotropic constitutive relations \cite{Guo,Gao}, the negative refractive index can occur within some frequency band and is not necessarily associated with simultaneously negative electric permittivity and negative magnetic permeability, being attributed to the chirality parameter introduced in the constitutive relations,
\begin{equation}  
\label{Guo-bi-isotropic-relations-1}
D^{i} =\varepsilon_{ij}E^{j} + i \xi_{c} B^{i} , \quad
H^{i} = \mu^{-1} B^{i} + i \xi_{c} E^{i} , 
\end{equation}
where $\varepsilon_{ij}$, $\mu$, and $\xi_{c}$ are the plasma electric permittivity tensor, the magnetic permeability, and the constant chirality parameter. Plasmas metamaterials have been investigated as new media endowed with interesting properties, such as negative refraction and nonlinearities \cite{Sakai, Sakai2}. {Rotating plasmas constitute a scenario in which the birefringence is enhanced, with special properties on the RP and attenuation \cite{Gueroult-3}. Therefore, plasmas provide a rich framework \textcolor{blue}{for} theoretical and experimental investigation on the optical behavior of chiral systems, as the one developed in this work.}

In this work, we are interested in examining the wave propagation in a magnetized cold plasma ruled by the MCFJ model, a chiral route distinct from the bi-isotropic/anisotropic electrodynamics of the relations (\ref{Guo-bi-isotropic-relations-1}).  We carry out our analysis considering the timelike Lorentz-violating background component, which plays the role of the chiral magnetic conductivity. {{Such a choice is analog to the coupling with a cold axion field, for which $\mathbf{\nabla }\theta={\bf{0}}$, as stated in \eqref{Maxwellaxion1}.} The refractive indices and dispersion relations are evaluated, representing altered collective electromagnetic modes. For the propagation along the magnetic field axis, in the low-frequency limit, there appear RCP and LCP helicons, due to the presence of the chiral fator $V_{0}$.} Optical effects, such as birefringence and dichroism, are examined, which could be useful to trace analogies with other material properties. We also find that the chiral conductivity yields negative refraction in specific frequency bands, amplifying the rotatory power and dichroism signals.

This paper is outlined as follows. In Sec.~\ref{themodel1}, we briefly review some aspects of the MCFJ model. In Sec.~\ref{usual-plasmas-section}, the main properties of propagation in usual cold magnetized plasmas are presented. {The dispersion relations, refractive indices, and helicons for cold plasma in chiral electrodynamics are adressed in Sec.~\ref{Wave-propagation-in-purely-time-like-background-case}.  In Sec.~\ref{sec5}, we discuss the case of propagation orthogonal to the magnetic field.} The optical effects are examined in Sec.~\ref{birefringence}. Finally, we summarize our results in Sec.~\ref{conclusion}.
	
\section{BASICS ON MCFJ ELECTRODYNAMICS \label{themodel1}}

The Carroll-Field-Jackiw model was proposed as a gauge invariant {\textit{CPT}-odd} electrodynamics constrained by birefringence data of distant galaxies \cite{CFJ}. It was later incorporated as the {\textit{CPT}-odd} sector of the SME \cite{Colladay}, and it has been investigated in several respects \cite{CFJ2, CFJ3}. In matter, it is described by the following Lagrangian density \cite{Pedroo}:\footnote{We use natural units $h=c=1$ and the Minkowski metric signature $g_{\mu\nu}=\mathrm{diag}\left( 1,-1,-1,-1\right)$.}
\begin{align}
\mathrm{{\mathcal{L}}}=-\frac{1}{4}G^{\mu\nu}F_{\mu\nu}-\frac{1}{4}%
\epsilon^{\mu\nu\alpha\beta}\left(  k_{AF}\right)  _{\mu}A_{\nu}F_{\alpha
	\beta}-A_{\mu}J^{\mu},\label{MCFJMATTER}%
\end{align}
yielding the MCFJ equation of motion,
\begin{equation}
\partial_{\rho}G^{\rho\kappa} + \epsilon^{\beta\kappa\mu\nu} \left(  k_{AF}\right)_{\beta} F_{\mu\nu} = J^{\kappa}\,.
\label{MotionEq1}
\end{equation}
Here, $\left(  k_{AF}\right)^{\mu}=\left( k_{AF}^{0}, {\bf{k}}_{AF}\right)$ is a constant 4-vector background responsible for the
Lorentz violation, and
\begin{align}
F_{\mu\nu}=\partial_{\mu}A_{\nu}-\partial_{\nu}A_{\mu}, \quad G^{\mu\nu}=\frac{1}{2}\chi^{\mu\nu\alpha\beta}F_{\alpha\beta},
\end{align} 
are the usual $U(1)$ vacuum and continuous matter field strength, respectively. 
The 4-rank tensor, $\chi^{\mu\nu\alpha\beta}$, describes the medium constitutive tensor \cite{refPOST}, whose components provide the electric and magnetic responses of the medium. Indeed, the electric permittivity and magnetic permeability tensor components are written as $\epsilon_{ij}\equiv \chi^{0ij0}$ and $\mu^{-1}_{lk}\equiv \frac{1}{4} \epsilon_{ijl}\chi^{ijmn}\epsilon_{mnk}$, respectively. For isotropic polarization and magnetization, it holds $\epsilon_{ij}=\epsilon  \delta_{ij}$ and $\mu^{-1}_{ij}=\mu^{-1} \delta_{ij}$, providing the usual isotropic constitutive relations,
\begin{align}
\mathbf{D}=\epsilon\mathbf{E}, \quad \mathbf{H}= \mu^{-1}\mathbf{B}. \label{CRusual}
\end{align}
A straightforward calculation from \eqref{MotionEq1} yields
\begin{align}
\nabla\cdot\mathbf{D}  &=J^{0}-\mathbf{k}_{AF}%
\cdot\mathbf{B}  ,\label{1}\\
\nabla\times\mathbf{H} -\frac{\partial\mathbf{D}
	  }{\partial t} &=\mathbf{J}+  k_{AF}^{0} \mathbf{B}+\mathbf{k}_{AF}\times
\mathbf{E} ,\label{2} 
\end{align}
where $G^{i0}=D^{i}$ and $G^{ij}=-\epsilon_{ijk}H^{k}$. The homogeneous Maxwell equations are given by
\begin{align}
\nabla\cdot\mathbf{B}  &=0, \quad \nabla\times\mathbf{E}  +\frac{\partial\mathbf{B}
	  }{\partial t} =\bf{0}. \label{4}%
\end{align} 
By using a plane-wave ansatz for the electromagnetic fields, the MCFJ equations (\ref{1}) (\ref{4}) {read,}
\begin{subequations}
\label{maxwell-equations-plane-wave-ansatz-1}
\begin{align}
i\mathbf{k}\cdot \mathbf{{D}} +\mathbf{k}_{AF}\cdot \mathbf{{B}}  &=J^{0}, \label{4.aac}\\
i\mathbf{k}\times \mathbf{{H}} +i\omega \mathbf{{D}} -
k_{AF}^{0}\mathbf{{B}} -\mathbf{k}_{AF} \times \mathbf{{E}} &=\mathbf{{J}}, \label{4.aab}\\
\mathbf{k}\cdot \mathbf{{B}} =0, \quad
\mathbf{k}\times \mathbf{{E}} -\omega 
\mathbf{{B}} &=\bf{0},  \label{4.aaa}
\end{align}
\end{subequations} 
where $\mathbf{k}$ is the wave vector and $\omega$ is the (angular) wave frequency.

In the presence of anisotropy, the permittivity and permeability are represented by rank 2 tensors, $\varepsilon_{ij}$ and $\mu_{ij}$, which may also depend on the frequency (for a dispersive medium). For an anisotropic medium, the constitutive relations (\ref{CRusual}) are replaced by \cite{refZANGWILL,refJACKSON},
\begin{equation}
D^{i}=\varepsilon_{ij}(\omega)E^{j}, \ \ \  B^{i}= \mu_{ij}(\omega) H^{j}. \label{CRAM}
\end{equation} 
For {nonmagnetic} media with isotropic magnetic permeability, it holds $\mu_{ij} (\omega)=\mu_{0}$, where $\mu_{0}$ is the vacuum permeability. Considering the constitutive relations (\ref{CRAM}), the modified Amp\`ere-Maxwell's law, Eq.(\ref{4.aab}), and Faraday's law, Eq.(\ref{4.aaa}), in the absence of sources, we obtain a modified wave equation for the electric field,
	\begin{equation}
	k^{i}\left(  k^{j}{E^{j}}\right)  -k^{2} {E}^{i} =-\omega^{2}\mu_{0}\bar{\varepsilon}_{ij}\left(  \omega\right)
	{E}^{j},\label{general}%
	\end{equation}
	where we define the extended permittivity tensor,
	{\begin{equation}
		\bar{\varepsilon}_{ij}( \omega ) = {\varepsilon}_{ij} (  \omega )+i \frac{k_{AF}^{0}}{\omega^2}\epsilon_{ikj}k^{k}+i \epsilon_{ikj}\frac{ k_{AF}^{k}}{\omega}.
		\label{permittivity1}
		\end{equation}}
Using the definition to the refractive index, $\mathbf{n}=\mathbf{k}/\omega$, the modified wave equation becomes 
	\begin{equation}
M_{ij} {E}^{j}=0,\label{general.1}				
\end{equation}
with $M_{ij}$ given by
\begin{equation}
M_{ij}=n^{2}\delta_{ij}-n_{i}n_{j}-\frac{\varepsilon_{ij}}{\varepsilon_{0}}- \frac{i}{\omega} \left(V_{0} \epsilon_{ikj}n^{k}+\epsilon_{ikj}V^{k} \right),
\end{equation} 
in which $\varepsilon_{0}$ is the vacuum electric permittivity, and
\begin{equation}
V_{0}=k_{AF}^{0}/\varepsilon_{0}, \quad V^{k}=k_{AF}^{k}/\varepsilon_{0}.
\end{equation} 
appear as the components of a redefined background, $V^{\mu}=\left(V_{0}, V^{i}\right)$. The nontrivial solutions for the electric field require a vanishing determinant of the matrix $M_{ij}$, $\det M_{ij}=0$,
which provides the dispersion relations that describe the wave propagation in the medium.

In this work, we will study plasma waves propagation for a chiral (parity-odd) medium, which means restraining our investigation to the case of a purely timelike Lorentz-violating background vector, $\left(  k_{AF}\right)^{\mu}=\left(
k_{AF}^{0}, \textbf{0}\right)$, {which also plays the role of chiral magnetic conductivity. This choice is physically meaningful since it represents cold dark matter, for which the space variation of the axion field can be neglected, $\mathbf{\nabla }\theta=\mathbf{k}_{AF}=0$.} In this scenario, the wave equation (\ref{general.1}) becomes

\begin{equation}
\left[n^{2}\delta_{ij}-n^{i}n^{j}-\frac{{\varepsilon}_{ij}%
}{\varepsilon_{0}}-i \frac{V_{0}}{\omega} \epsilon_{ikj}n^{k} \right] {E}^{j}=0.
\label{EWE2}
\end{equation}

\section{The usual magnetized cold plasma}\label{usual-plasmas-section}

In this work we will adopt the fluid theory approach in the cold plasma limit \cite{Gurnett,STURROK,Boyd,Stix,Bittencourt}:
\begin{gather}
\frac{\partial n}{\partial t}+\mathbf{\nabla}\cdot\left(  n\mathbf{u}\right)
=0,\label{7}\\
\frac{\partial\mathbf{u}}{\partial t}+\mathbf{u}\cdot\mathbf{\nabla u}%
=\frac{q}{m}\left(  \mathbf{E}+\mathbf{u}\times\mathbf{B}_{0}\right)
,\label{7.a}%
\end{gather}
 where $n$ is the electron number density, $\mathbf{u}$ is the electron fluid
	velocity field, $q$ and $m$ are the electron charge and mass, respectively, and
	$\mathbf{B}_{0}$ is the equilibrium magnetic field. For simplicity, the ions are supposed to be infinitely massive, which is appropriate for high-frequency waves. Furthermore, thermal and collisional effects are also disregarded.
The linearized version of the magnetized cold plasmas \cite{Boyd} consider fluctuations around average quantities, $n_{0}$ and $\mathbf{B}_{0}$, which are constant in time and space. Thus, the plasma quantities read
\begin{subequations}
\label{cold-plasma-fluctuations-from-equilibrium-1}
\begin{align}
	 n&=n_{0}+\delta n,\\
	  \mathbf{u}&=\delta\mathbf{u},\\ \mathbf{E}&=\delta\mathbf{E} \\
	\mathbf{B}&=\mathbf{B}_{0}+\delta\mathbf{B},
\end{align}
\end{subequations}
with $\delta n$,
$\delta\mathbf{u}$, $\delta\mathbf{E}$ and $\delta\mathbf{B}$ being {first-order} plane wave magnitude perturbations. Following the usual procedure \cite{Gurnett, STURROK, Stix, Bittencourt}, assuming $\mathbf{B}%
_{0}=B_{0}\hat{z}$, we write the corresponding dielectric tensor,
\begin{equation}
\varepsilon_{ij}  (\omega)=\varepsilon_{0}%
\begin{bmatrix}
S & -iD & 0\\
iD & S & 0\\
0 & 0 & P
\end{bmatrix}
,\label{7.1}%
\end{equation}
where
\begin{equation}
S=1-\frac{\omega_{p}^{2}}{\left(  \omega^{2}-\omega_{c}^{2}\right)
},\   D=\frac{\omega_{c}\omega_{p}^{2}}{\omega\left(  \omega^{2}-\omega
	_{c}^{2}\right)  },\  P=1-\frac{\omega_{p}^{2}}{\omega^{2}}, \label{def}
\end{equation}
and
\begin{equation}
\omega_{p}=\frac{  n_{0}q^{2}}{m\epsilon_{0}}, \quad  \omega_{c}=\frac{|q| B_{0}}{m} , \label{plasmas-characteristic-frequencies-1}
\end{equation}
are the plasma and cyclotron frequencies, respectively. 

{In cold magnetized plasmas, it is usual to investigate modes that propagate parallel and perpendicular to the magnetic field. For longitudinal propagation to the magnetic field, $\mathbf{k}\parallel\mathbf{B}_{0}$,} two distinct refractive indices are obtained, 
\begin{equation}
n_{\pm}=\sqrt{1- \frac{\omega_{p}^{2}} {\omega\left(
	\omega\pm\omega_{c}\right)}} , \label{nusual2}
\end{equation}
which provide {left-circularly polarized and right-circularly polarized} modes, respectively,

\begin{equation}
\mathbf{{E}}_{LCP}=\frac{i}{\sqrt{2}}%
\begin{bmatrix}
1 \\ 
i%
\end{bmatrix}, \quad
\mathbf{{E}}_{RCP}=\frac{i}{\sqrt{2}}%
\begin{bmatrix}
1 \\ 
-i%
\end{bmatrix},\label{ERCP}
\end{equation} 
for the propagating modes associated to $n_{\pm}$, respectively. This is the standard result of wave propagation in the usual magnetized cold plasma.
We recall that a cutoff happens whenever the refractive index, $n$, goes to zero. On the other hand, a resonance occurs if $n$ tends to infinity.
From the indices (\ref{nusual2}), we obtain the following cutoff frequencies:	
\begin{align}
\omega_{\pm} &=\frac{1}{2}\left(  \sqrt{\omega_{c}^{2}+4\omega_{p}^{2}}\mp\omega_{c}\right),\label{r1}
\end{align}
where $\omega_{\pm}$ is related to $n_{\pm}$, respectively.

{As for perpendicular propagation to the magnetic field, $\mathbf{k}\perp\mathbf{B}_{0}$,  $\mathbf{k}=(k_{x},k_{y},0)$, two refractive indices are obtained. The one corresponding to the transversal mode, with $\delta\mathbf{E}=(0,0,\delta E_{z})\perp\mathbf{k}$,  is
{	\begin{equation}
	n_{T}^{2}=P, \label{nusualort1}
	\end{equation} }
while the \textit{extraordinary} mode \cite{Boyd}, 
	{\begin{equation}
		n_{O}^{2}=\frac{\left(S+D\right)\left(S-D\right)}{S}, \label{nusualort2}
		\end{equation}}
is longitudinal, that is, $\delta\mathbf{E}=(\delta E_{x},\delta E_{y},0)$. The parameters $P$, $S$ and $D$ are given in \eqref{def}. The refractive index $n_{T}$ provides a linearly polarized mode, whereas $n_{O}$, in general, is related to an elliptically polarized mode.}

A very usual effect in magnetized plasmas is the circular birefringence\footnote{In plasmas, the birefringence is usually a consequence of the Faraday effect, occurring due to the presence of the external field  $\mathbf{B}_{0}$, which generates distinct phase velocities for the propagating modes \cite{Porter}.}, which causes the rotation of the plane of polarization of a linearly polarized wave that propagates within the medium. Thus the linearly polarized wave emerges from the medium with an electric field whose polarization is rotated relative to its initial linear configuration. Such a phenomenon can be properly explained by decomposing the initial wave into two circularly polarized waves (RCP and LCP) that travel with different phase velocities. In this case, the rotation angle of the electric field can be expressed as the difference between the refractive indices associated with the RCP and LCP waves \cite{Fowles, Condon}:
\begin{equation}
\theta =\frac{\pi L}{\lambda_{0} }%
\left( \mathrm{Re}\left[n_{RCP}\right]-\mathrm{Re}\left[n_{LCP}\right]\right) ,
\end{equation}%
where $\lambda_{0} $ is the vacuum wavelength of the incident wave. The rotation power $\delta=\theta/L$ (phase difference per unit length), is given as
\begin{equation}
\delta =-\frac{\omega }{2 }\left( \mathrm{Re}\left[n_{LCP}\right]-\mathrm{Re}\left[n_{RCP}\right]\right).  \label{powerROT}
\end{equation}

{For parallel propagation in a cold magnetized plasma, $\mathbf{k}\parallel\mathbf{B}_{0}$,} the refractive indices (\ref{nusual2}) provide the following rotatory power:
\begin{equation}
\delta =-\frac{\omega }{2}\mathrm{Re} \left( \sqrt{1-\frac{\omega_{p}^{2}}{\omega\left(\omega+\omega_{c}\right)}}- \sqrt{1-\frac{\omega_{p}^{2}}{\omega\left(\omega-\omega_{c}\right)}} \right). \label{rpcasousual1}
\end{equation}

\begin{figure}[h]
	\centering
	\includegraphics[scale=0.65]{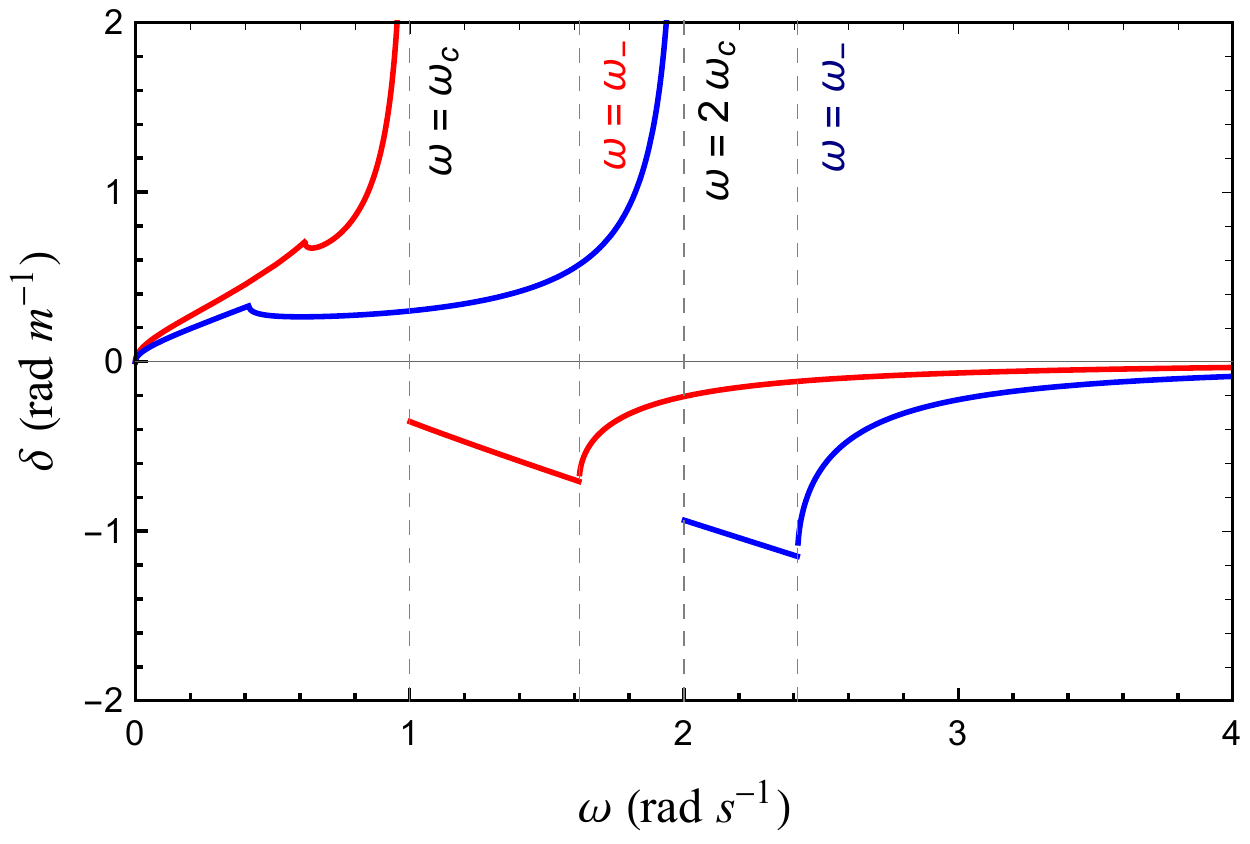}  \caption{Rotatory power (\ref{rpcasousual1}) in terms of $\omega$.  Here, $\omega_{c}=\omega_{p}$ (red line) and $\omega_{c}=2\omega_{p}$ (blue line), with the choice $\omega_{c}=1$~$\mathrm{rad}$~$s^{-1}$.}
	\label{rp1}%
\end{figure}

The behavior of the RP (\ref{rpcasousual1}) in terms of the frequency $\omega$ is depicted in Fig.~\ref{rp1}. One notices that there is a divergence at $\omega_{c}$, being positive for $\omega<\omega_{c}$ and negative for $\omega>\omega_{c}$. It tends to zero at the high-frequency limit $\omega>>\left(\omega _{p},\omega _{c}\right)$, where it decays as 
\begin{equation}
\delta \approx -\frac{\omega _{p}^{2}\omega _{c}}{2\omega ^{2}}.
\label{rotFaraday}
\end{equation}
Associated with the imaginary part of the refractive index, one can also examine dichroism, an optical effect that occurs when circularly polarized waves are absorbed by the medium at different degrees \cite{Pedroo,Hecht,Wagniere}. Thus dichroism coefficient refers to the difference in absorption of LCP and RCP waves, being given by: 
\begin{equation}
	\delta_{d} =-\frac{\omega }{2 }\left( \mathrm{Im}[n_{LCP}]-\mathrm{Im}[n_{RCP}]\right).  \label{dicro}
\end{equation}
which, for the refractive indices (\ref{nusual2}), implies
	\begin{equation}
		\delta_{d} =-\frac{\omega }{2} \mathrm{Im} \left( \sqrt{1-\frac{\omega_{p}^{2}}{\omega\left(\omega+\omega_{c}\right)}}- \sqrt{1-\frac{\omega_{p}^{2}}{\omega\left(\omega-\omega_{c}\right)}} \right).\label{dicrousualcase}
	\end{equation}

\begin{figure}[h]
	\centering
	\includegraphics[scale=0.65]{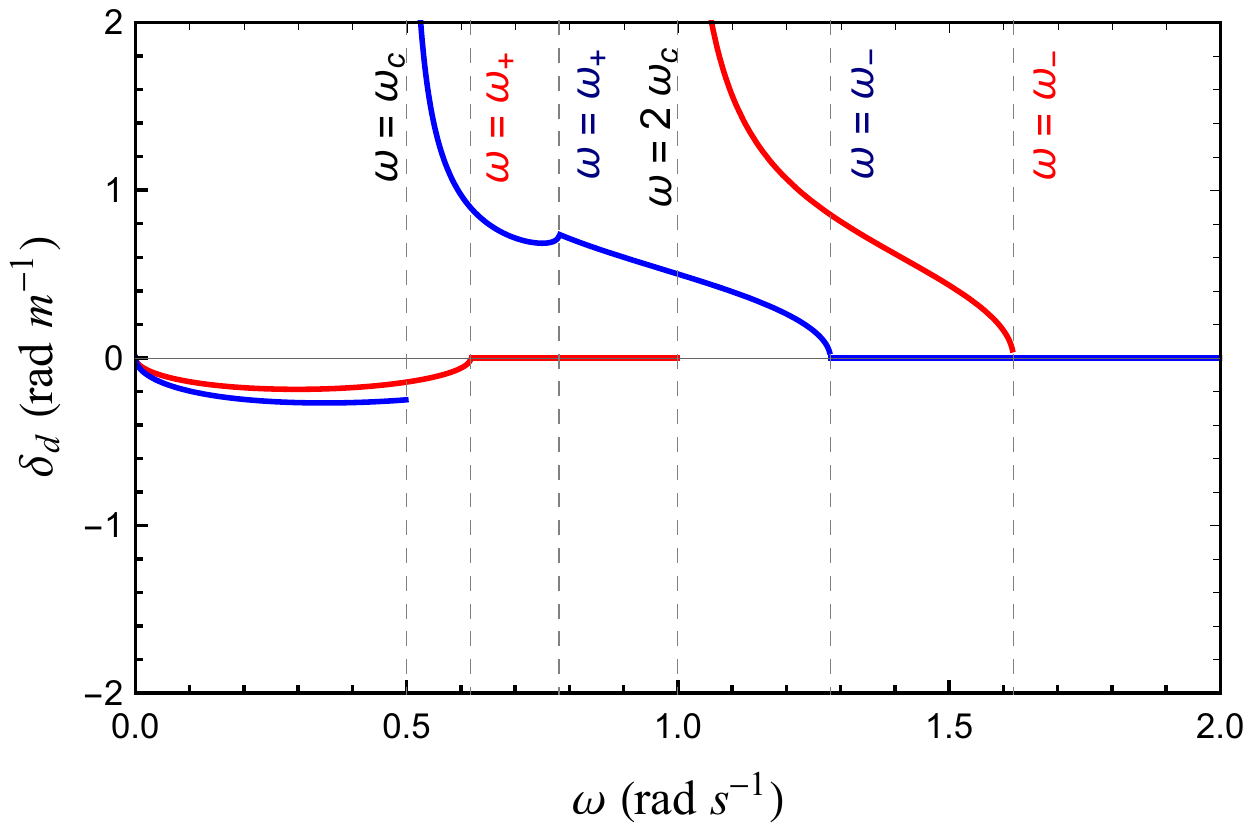}  \caption{Dichroism coefficient (\ref{dicrousualcase}) in terms of $\omega$.  Here, $\omega_{c}=\omega_{p}$ (red line) and $\omega_{c}=\omega_{p}/2$ (blue line), with $\omega_{c}=1$~$\mathrm{rad}$~$s^{-1}$.}
	\label{dicrousual}%
\end{figure}

	Such a quantity is plotted in Fig. \ref{dicrousual}, which shows singularity at the cyclotron frequency $\omega_{c}$. For $\omega_{c}=\omega_{p}$ (red curve), the dichroism coefficient (\ref{dicrousualcase}) is negative for $\omega<\omega_{+}^{red}$, positive for $2\omega_{c}<\omega<\omega_{-}^{red}$ and null for other frequencies. The case for $\omega_{c}=\omega_{p}/2$ (blue curve) differs in the fact that $\omega_{+}^{blue}$ is greater than $\omega_{c}$, showing that (\ref{dicrousualcase}) is now negative for $\omega<\omega_{c}$.

{\section{Wave propagation in chiral plasma along the magnetic field}
\label{Wave-propagation-in-purely-time-like-background-case} }


{Collective modes of chiral systems were examined in the context of the Weyl materials by means of the chiral kinetic theory \cite{Gorbar1}. Using a similar formalism, low-energy collective modes (pseudomagnetic helicons) were predicted for Dirac and Weyl matter \cite{Gorbar2}. In this section, we derive the collective electromagnetic modes of the cold chiral plasma ruled by the permittivity (\ref{permittivity1}). In this sense, we start from the wave equation (\ref{EWE2}) and use the expression of the magnetized} plasma dielectric permittivity, given in \eqref{7.1}, {obtaining} a linear homogeneous system,

\begin{widetext}
\begin{equation}
\begin{bmatrix}
n^{2}-n_{x}^{2}-S & iD-n_{x}n_{y}+i\left(V_{0}/\omega\right)n_{z} &
-n_{x}n_{z}-i\left(V_{0}/\omega\right)n_{y}\\
-iD-n_{x}n_{y}-i\left(V_{0}/\omega\right)n_{z} & n^{2}-n_{y}^{2}-S & -n_{y}n_{z}+i\left(V_{0}/\omega\right)n_{x}\\
-n_{x}n_{z}+i\left(V_{0}/\omega\right)n_{y} & -n_{y}n_{z}-i\left(V_{0}/\omega\right)n_{x} & n^{2}-n_{z}^{2} -P
\end{bmatrix}
\begin{bmatrix}
\delta E_{x}\\
\delta E_{y}\\
\delta E_{z}%
\end{bmatrix}
=0. \label{timelike2}
\end{equation}
\end{widetext}


Let us consider, for simplicity, the case the refractive index is parallel to the magnetic field, $\mathbf{n}=n\hat{z}$, such that one obtains
\begin{equation}
\begin{bmatrix}
n^{2}-S & iD+i\left(V_{0}/\omega\right)n &
0\\
-iD-i\left(V_{0}/\omega\right)n & n^{2}-S & 0\\
0 & 0 & -P
\end{bmatrix}
\begin{bmatrix}
\delta E_{x}\\
\delta E_{y}\\
\delta E_{z}%
\end{bmatrix}
=0, \label{timelike}
\end{equation}
for which $\mathrm{det}[M_{ij}] =0$ provides the dispersion relations
\begin{equation}
P \left(  \omega^{2}
	\left(  n^{2}-S\right)  ^{2}-( \omega D +n V_{0})^{2}\right)=0.
	\label{DR1A}
\end{equation}
Longitudinal waves, $\mathbf{n}\parallel
\delta\mathbf{E}$ or $\delta\mathbf{E}=(0,0,\delta E_{z})$, may emerge, when $P=0$, with {nonpropagating} vibration at the
plasma frequency, $\omega=\omega_{p}$. {Under an electromagnetic perspective, this longitudinal oscillation is a plasmon. In a solid-state context, the collective mode of electrons vibrating (longitudinally) under the action of the electromagnetic field is also called plasmons.}

 For transverse waves,   $\mathbf{n}\perp\delta\mathbf{E}$ or $\delta\mathbf{E}=(\delta E_{x},\delta E_{y},0)$, the dispersion relation (\ref{DR1A}) simplifies as
\begin{equation}
\left(n^2-S\right)^2-\left(D +n \left(V_{0}/\omega\right) \right)^2=0,
\end{equation}
also written as a fourth-order equation in $n$,
\begin{equation}
n^4-\left(2S+ \left(V_{0}/\omega\right)^2\right)n^2-2D \left(V_{0}/\omega\right)n+\left(S^{2}-D^2\right)=0. 
\label{DR1B}
\end{equation}
Taking into account the relations (\ref{def}), the dispersion relation (\ref{DR1B}) {provides the following refractive indices for electromagnetic modes} of the model:
\begin{align}
n_{R,M} &= - \frac{V_{0}}{2\omega} \pm \sqrt{ 1 + \left( \frac{V_{0}}{2\omega}\right)^{2} - \frac{\omega_{p}^{2}}{\omega (\omega - \omega_{c})}} , \label{n-R-M-indices-1} \\
n_{L,E} &= \frac{ V_{0}}{2\omega} \pm \sqrt{ 1 + \left( \frac{V_{0}}{2\omega}\right)^{2} - \frac{\omega_{p}^{2}}{\omega (\omega + \omega_{c}) }}. \label{n-L-E-indices-1}
\end{align}

In general, the indices $n_{R}, n_{L}, n_{E}, n_{M}$ may be real, imaginary, or complex (presenting both pieces) at some frequency ranges. As well-known, the real part is associated with propagation, while the complex piece is concerned with absorption. Furthermore, these indices may have positive or negative real pieces. The indices $n_{L}$ and $n_{M}$ are always positive and negative, respectively, the latter one being a negative refractive index. On the other hand, the indices $n_{R}$ and $n_{E}$ can be positive or negative, depending on the frequency zone examined, in such a way the associated modes can manifest negative refraction behavior (in a suitable frequency band).

The propagating modes associated with the refractive indices in \eqref{n-R-M-indices-1}) and \eqref{n-L-E-indices-1} are obtained by inserting each one in Eq. (\ref{timelike}) and carrying out the corresponding eigenvector (with a null eigenvalue).  The emerging electric field are the $\mathbf{{E}}_{LCP}$ and $\mathbf{{E}}_{RCP}$, given in Eq. (\ref{ERCP}), where $n_R$, $n_M$ are associated with the RCP mode, and $n_L$, $n_{E}$ are related to the LCP mode,
\begin{equation}
n_{L}, n_{E}  \   \mapsto  \ \mathbf{{E}}_{LCP}=\frac{i}{\sqrt{2}}%
\begin{bmatrix}
1 \\ 
i%
\end{bmatrix}, \label{ELCP2}
\end{equation}
\begin{equation}
n_{R}, n_{M} \   \mapsto  \ \mathbf{{E}}_{RCP}=\frac{i}{\sqrt{2}}%
\begin{bmatrix}
1 \\ 
-i%
\end{bmatrix}.\label{ERCP2}
\end{equation} 

From the indices $n_R$, $n_{E}$, given by Eqs. (\ref{n-R-M-indices-1}) and (\ref{n-L-E-indices-1}), we obtain the same cutoff frequencies (\ref{r1}) of the standard case: in fact, $\omega_{-}$ is related to the refractive index $n_{R}$, and $\omega_{+}$ is associated with the refractive index $n_{E}$. In contrast, the refractive indices $n_{L}$ and  $n_{M}$ have no real root. The behavior of the refractive indices in Eqs. (\ref{n-R-M-indices-1}) and (\ref{n-L-E-indices-1}) will be examined in the following.

\subsection{About the index $n_{R}$ \label{secNR}}

\begin{figure}[h]
	\centering
	\includegraphics[scale=0.72]{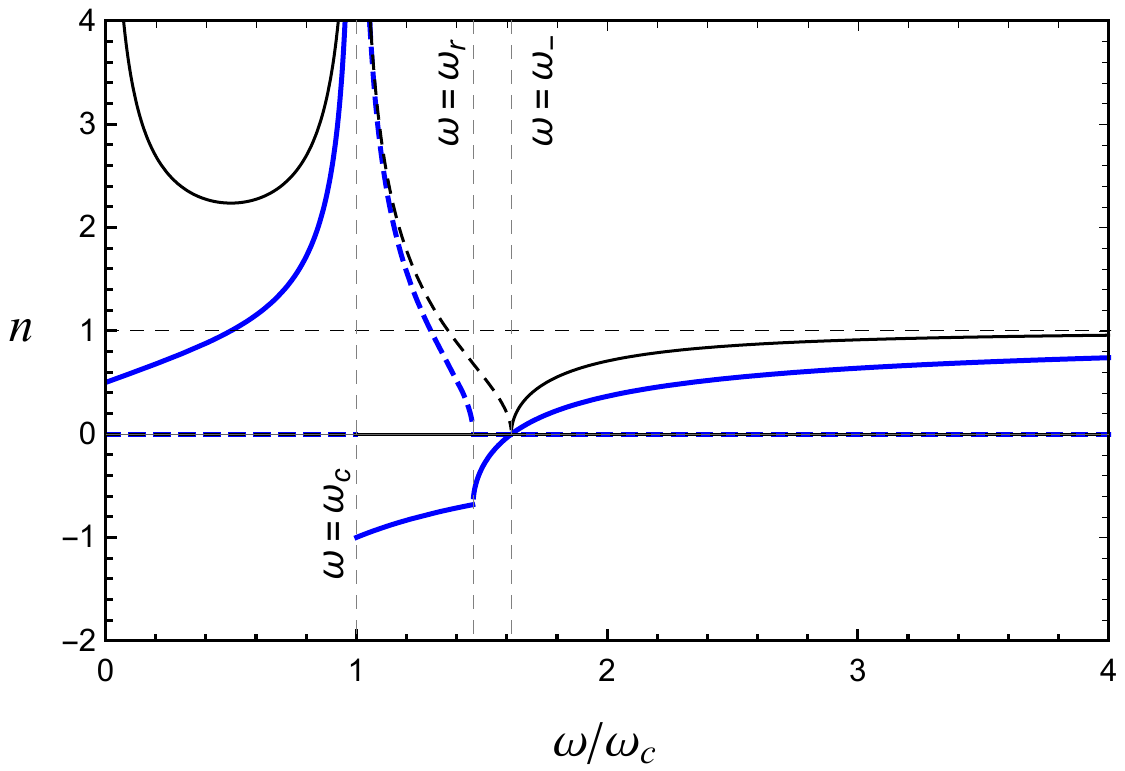}  \caption{Index of refraction 
		$n_{R}$ in terms of the frequency $\omega$. The dashed blue (black) line corresponds to the imaginary piece of $n_{R}$ ($n_{-}$), while the solid blue (black) line represents the real piece of $n_{R}$ ($n_{-}$). Here $\omega_{c}=\omega_{p}$, $V_{0}=2\omega_{p}$, and $\omega_{c}=1$~$\mathrm{rad}$~$s^{-1}$.}
	\label{nRfig}%
\end{figure}

We initiate discussing some properties of the index $n_{R}$. The behavior of $n_{R}$ in terms of the dimensionless parameter $\omega/\omega_{c}$ is illustrated in Fig.~\ref{nRfig}, which displays the real imaginary pieces of the refractive index $n_{R}$.  We point out:

\begin{enumerate}
	[label=(\roman*)]
	
	\item It takes on a finite value 
	when $\omega\rightarrow 0$, given by 
	\begin{equation}
		 n_{R}\left( 0\right) =\frac{1}{V
			_{0}}\left( \frac{\omega _{p}^{2}}{\omega _{c}}\right),
		\end{equation}
{describing a modified helicon in relation to the standard scenario, where the usual magnetized index $n_{-}$ goes to infinity} near the origin.
	
	\item For $0<\omega<\omega _{c}$, $n_{R}$ is positive since the square root in (\ref{n-R-M-indices-1}) is real, positive, and larger than the negative piece before it. Such a positivity also holds for the usual index $n_{-}$. See the black line in this frequency zone in Fig.~\ref{nRfig}.

	\item For $\omega\rightarrow\omega _{c}$, $n_{R}\rightarrow\infty$,
	and there occurs a resonance at the cyclotron frequency.
	
	\item For $\omega_{c}<\omega<\omega_{r}$, there appears a negative refractive index zone with absorption, where $\mathrm{Re}[n_{R}]<0$ and $\mathrm{Im}[n_{R}]\ne0$, as shown in Fig.~\ref{nRfig}. The frequency $\omega_{r}$ is the root of the radicand in Eq. (\ref{n-R-M-indices-1}),
		\begin{equation}
		R_{-}\left(  \omega\right)  =1+\frac{V  _{0}^{2}}{4\omega^{2}%
		}- \frac{\omega_{p}^{2}}{\omega\left(  \omega-\omega_{c}\right)  }, \label{radnR}
		\end{equation}
		which yields a cubic equation in $\omega$.

	\item For $\omega _{r}<\omega<\omega_{-}  $, one finds a negative refractive index zone without absorption, that is, $\mathrm{Re}[n_{R}]<0$ and $\mathrm{Im}[n_{R}]=0$.
	
	\item For $\omega>\omega_{-}$, the quantity $n_{R}$ is always positive, corresponding to a propagating zone, with $n_{R} \rightarrow1$ in the high-frequency limit.
\end{enumerate}

The frequency zone in which $\mathrm{Im}[n_{R}]\neq0$, that is, $\omega_{c}<\omega<\omega_{r}$, corresponds to the absorption zone for the metamaterial (negative refractive index) RCP wave, as already mentioned before. The frequency ranges in which $\mathrm{Im}[n_{R}]=0$ define the propagation zone for the RCP wave.

\subsection{About the index $n_{L}$ \label{secNL}}

The index $n_{L}$, given in Eq. (\ref{n-L-E-indices-1}), has no real root, presenting the following features:

\begin{enumerate}
	[label=(\roman*)]
	
	\item For $\omega\rightarrow0$, $n_{L}$ $\rightarrow+\infty$. Then, the presence of the term $V_{0}$ turns the refractive index real and positively divergent at the origin, differing from the usual index $n_{+}$ behavior, see \eqref{nusual2}, which is complex and divergent, $\mathrm{Im}[n_{+}]\rightarrow\infty$, at the origin. {As it will be clear in \eqref{helicons-16} of Sec. E, this index also supports helicons, or more specifically, LCP helicons.}
		
	\item For $\omega>0$, it is necessary to analyze the radicand in Eq. (\ref{n-L-E-indices-1}),
	\begin{equation}
	R_{+}\left(  \omega\right)  =1+\frac{V  _{0}^{2}}{4\omega^{2}%
	}- \frac{\omega_{p}^{2}}{\omega\left(  \omega+\omega_{c}\right)  }, \label{rad}
	\end{equation}	
	since it can be positive or negative, which determines the absence or presence of an absorption zone, respectively. Note that for $\omega>\omega_{+}$ the term $1-\omega_{p}^{2}/\omega\left(
	\omega+\omega_{c}\right)  $ is greater than zero ($\omega_{+}$ is the root of such a term), such that $R_{+}$ is positive. Therefore, the possibility of $R_{+}$ being negative occurs only in the range $0<\omega<\omega_{+}$, for which the term $1-\omega_{p}
		^{2}/\omega\left(  \omega+\omega_{c}\right)$ is less than zero. Hence, this positivity for $R_{+}$ is stated by the condition,
	\begin{equation}
	\frac{V_{0}^{2}}{4\omega^{2}}>\left\vert 1-\frac{4\omega
		_{p}^{2}}{\omega(\omega+\omega_{c})  }\right\vert _{\omega<\omega_{+}%
	},\label{cond}%
	\end{equation}
for which $R_{+}$ is always positive and the refractive index $n_{L}$ is real for any $\omega>0$. This corresponds to a propagating mode for the entire frequency domain. The behavior of $n_{L}$ in terms of the dimensionless parameter $\omega/\omega_{c}$, considering the condition (\ref{cond}), that is, $R_{+}>0$, is shown in Fig.~\ref{nLfig}.

	\item On the other hand, for
	\begin{equation}
	\frac{V_{0}^{2}}{4\omega^{2}}<\left\vert 1-\frac{4\omega
	_{p}^{2}}{\omega(\omega+\omega_{c})  }\right\vert _{\omega<\omega_{+}},\label{cond2}
	\end{equation}
	one has $R_{+}<0$ and $n_{L}$ becomes complex,  $\mathrm{Im}[n_{L}]\neq 0$, determining the opening of an absorption zone located within the interval $\omega_{i}<\omega<\omega_{f}$, as shown in Fig.~\ref{nLgap}. The frequencies $\omega_{i}$ and $\omega_{f}$ are positive and real roots of $R_{+}$, a cubic equation in the frequency.

	\begin{figure}[h]
	\centering
	\includegraphics[scale=0.72]{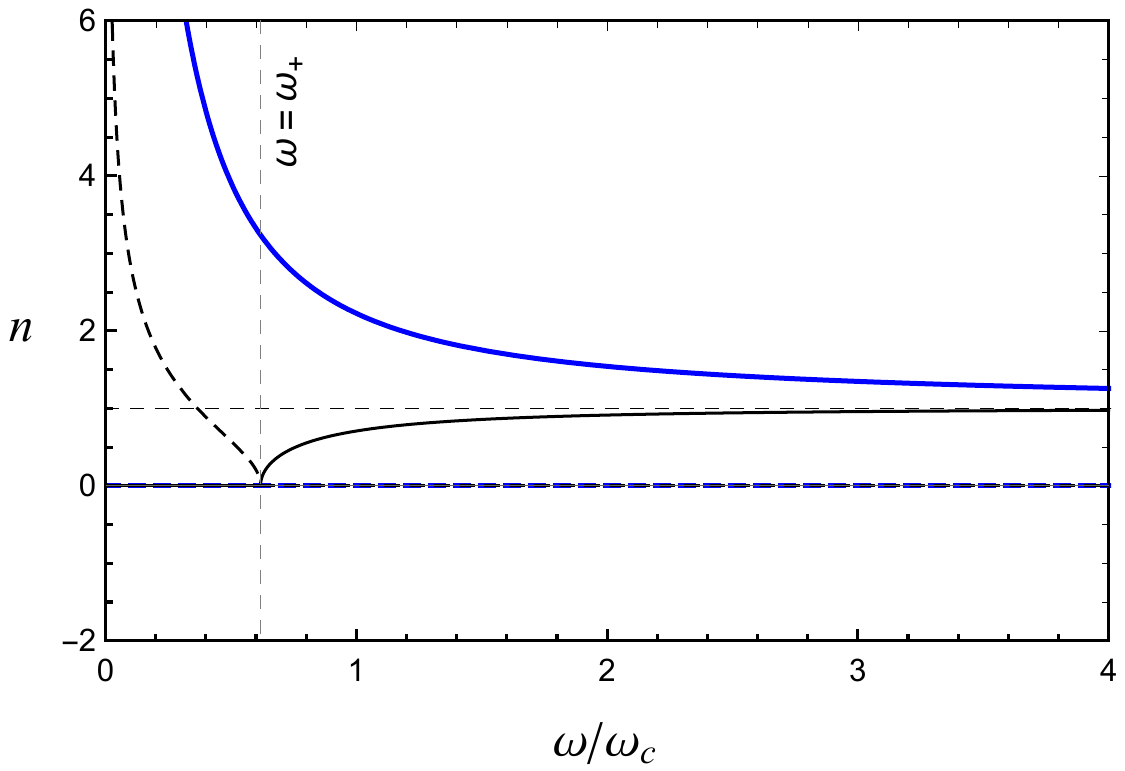}
	\caption{Refractive index  $n_{L}$ (blue lines) for the condition (\ref{cond}), $R_{+}>0$. Refractive index $n_{+}$ (black lines) of \eqref{nusual2}. The dashed (solid) lines correspond to the imaginary (real) pieces of $n_{L}$ and $n_{+}$. Here $\omega_{c}=\omega_{p}$, $V_{0}=2\omega_{p}$, and $\omega_{c}=1$~$\mathrm{rad}$~$s^{-1}$.}
	\label{nLfig}	
\end{figure}
\begin{figure}[h]		
			\centering
			\includegraphics[scale=0.72]{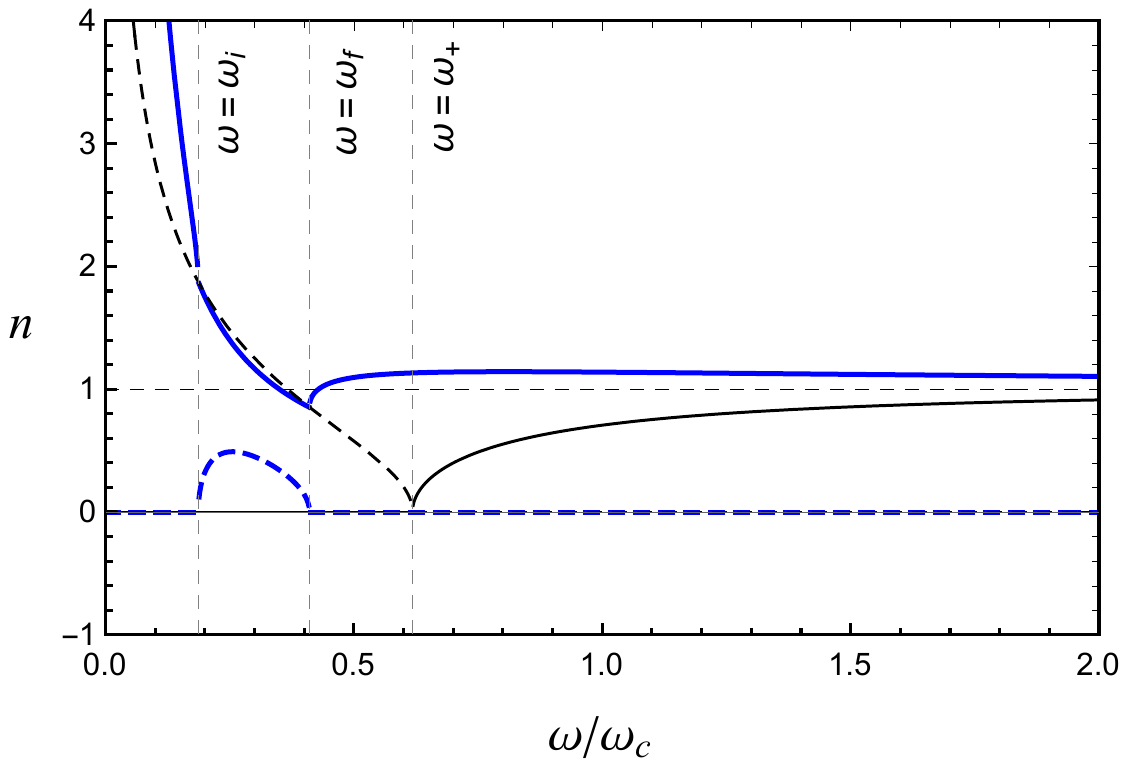}
			\caption{Refractive index  $n_{L}$ (blue lines) for the condition (\ref{cond2}), $R_{+}<0$. Refractive index $n_{+}$ (black lines) of \eqref{nusual2}. The dashed (solid) lines correspond to the imaginary (real) pieces of $n_{L}$ and $n_{+}$. Here $\omega_{c}=\omega_{p}$, $V_{0}=0.7\omega_{p}$, and $\omega_{c}=1$~$\mathrm{rad}$~$s^{-1}$.}
			\label{nLgap}
	\end{figure}

\end{enumerate}

\subsection{About the index $n_{E}$ \label{secNE}}

\begin{figure}[h]
	\centering
	\includegraphics[scale=0.72]{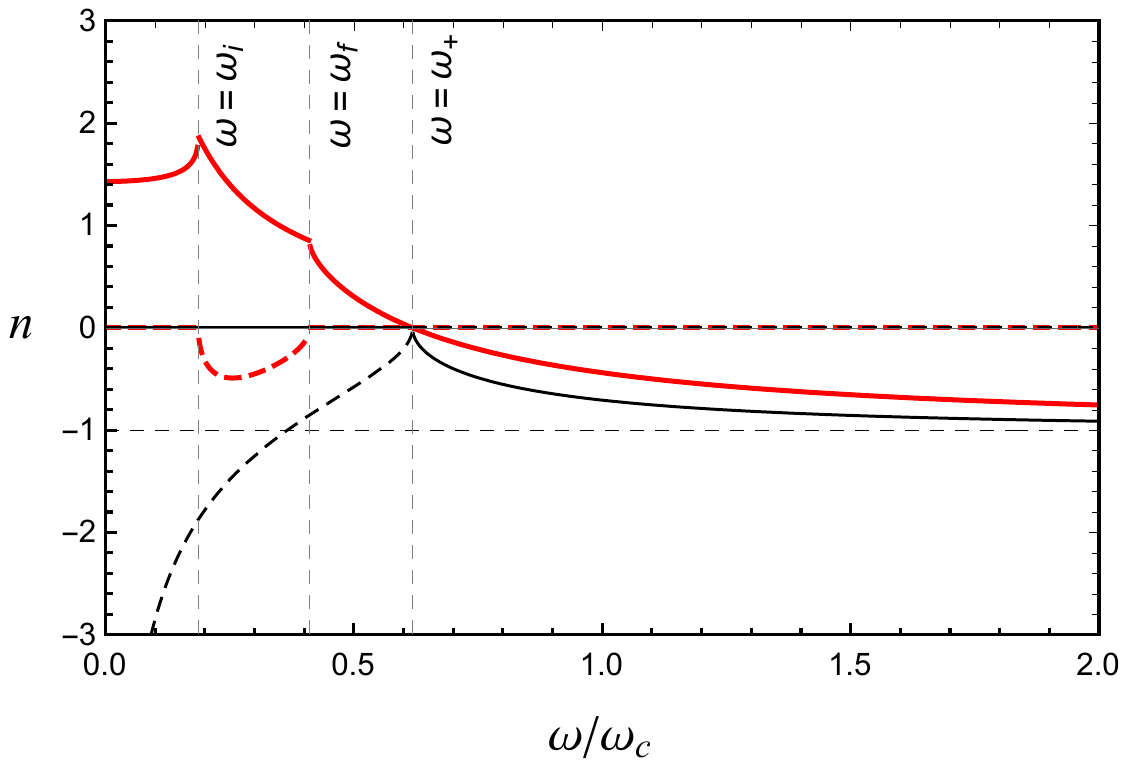}
	\caption{Red line: plot of the index $n_{E}$ for the condition (\ref{cond2}), $R_{+}<0$. Black line: plot of the index $-n_{+}$ of \eqref{nusual2}. Dashed (solid) lines represent the imaginary (real) pieces of $n_{E}$ and $-n_{+}$. Here, we have set $\omega_{c}=\omega_{p}$, $V_{0}=0.7\omega_{p}$, and $\omega_{c}=1$~$\mathrm{rad}$~$s^{-1}$.}
	\label{nEgap}
\end{figure}

The quantity $n_{E}$ is a refractive index that only exists as a positive quantity due to the presence of the chiral Lorentz-violating term. In the case we set $V_{0}=0$, the second relation in Eq. (\ref{n-L-E-indices-1}) yields $\mathrm{Re}[n_{E}]<0$ (negative index of refraction). For $V_{0}\neq0$, the index $n_{E}$ presents a small positivity range, $\mathrm{Re}[n_{E}]>0$, which provides propagation for the associated LCP wave. We present below some aspects of $n_{E}$: 
\begin{enumerate}
	[label=(\roman*)]
	
	\item For $\omega\rightarrow0$, the index $n_{E}$ tends to a finite value at origin,
	\begin{equation}
	n_{E}\left( 0\right) =\frac{1}{V  _{0}}\left[
	\frac{\omega_{p}^{2}}{\omega_{c}}\right]  ,
	\end{equation}
which is inversely proportional to the magnitude of the chiral factor, $V _{0}$. {Such a distinct low-frequency behavior also implies a LCP helicon, inexistent in the usual case.}
	
	\item Since the radicand of $n_{E}$ is the same one of $n_{L}$, see \eqref{n-L-E-indices-1}, it holds here the same procedure applied for $n_{L}$. For values of $V_{0}$ that satisfy the condition (\ref{cond}), $R_{+}>0$, $n_{E}$ is always real, $\mathrm{Im}[n_{E}]=0$, being positive within the interval $ 0<\omega<\omega_{+}$, and negative for $ \omega>\omega_{+}$, since $\sqrt{R_{+}}>V _{0}/2\omega$ at this range. The real and imaginary parts of $n_{E}$ are represented in Fig.~\ref{nEfig}.

	\begin{figure}[h]
		
			\centering
			\includegraphics[scale=0.72]{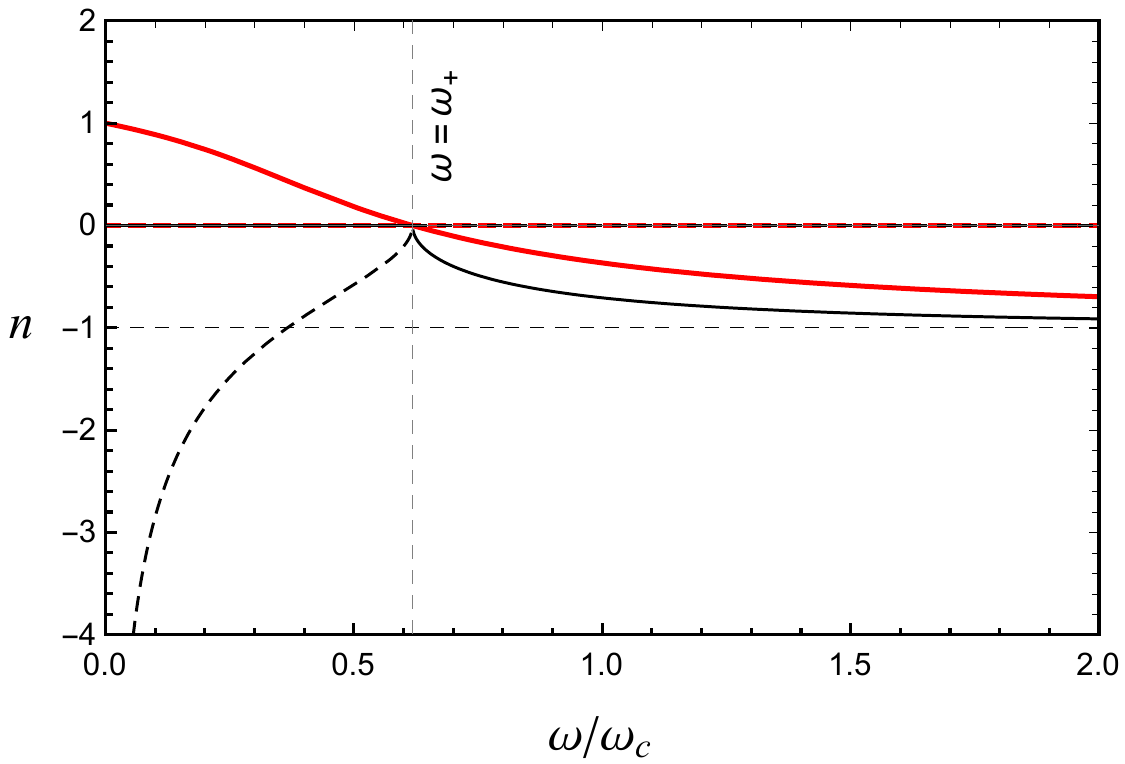}
			\caption{Red line: plot of the index $n_{E}$ for the condition (\ref{cond}), $R_{+}>0$. Black line: plot of the index $-n_{+}$ of \eqref{nusual2}. Dashed (solid) lines represent the imaginary (real) pieces of $n_{E}$ and $-n_{+}$. Here, we have used $\omega_{c}=\omega_{p}$ and $V_{0}=2\omega_{p}$, with the choice $\omega_{c}=1$~$\mathrm{rad}$~$s^{-1}$.}
			\label{nEfig}

	\end{figure}

	\item Considering the condition (\ref{cond2}), $n_{E}$ becomes complex and exhibits an absorption zone, $\mathrm{Im}[n_{E}]\neq0$, in the interval $\omega_{i}<\omega<\omega_{f}$, with $\omega_{i},\omega_{f}<\omega_{+}$, as shown in Fig.~\ref{nEgap}. Such a figure depicts the real and imaginary pieces of $n_{E}$ {[under the condition (\ref{cond2})]}.
\end{enumerate}

\subsection{About the index $n_{M}$ \label{secNM}}

\begin{figure}[h]		
	\centering
	\includegraphics[scale=0.72]{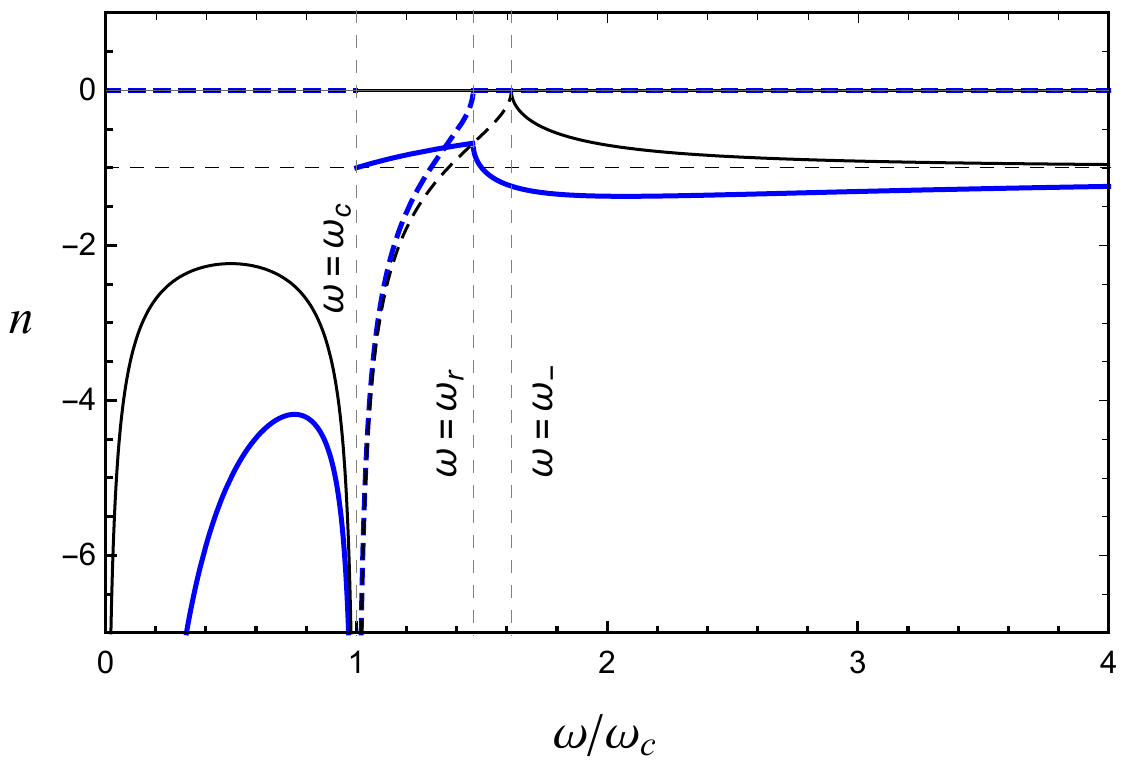}
	\caption{Blue line: plot of the index $n_{M}$. Black line: plot of the real piece of $-n_{-}$ of \eqref{nusual2}. Dashed (solid) lines represent the imaginary (real) pieces of $n_{M}$ and $-n_{-}$. Here, we have used: $\omega_{c}=\omega_{p}$, $V_{0}=2\omega_{p}$, and $\omega_{c}=1$~$\mathrm{rad}$~$s^{-1}$.}
	\label{nMeta}
\end{figure}

The additional index $n_{M}$, given in Eq. (\ref{n-R-M-indices-1}), is always negative (negative refraction) and has no real root. The behavior of $n_{M}$ in terms of the dimensionless parameter $\omega/\omega_{c}$ is shown in Fig.~\ref{nMeta}. We notice the following features:
\begin{enumerate}
	[label=(\roman*)]
	
	\item For $0<\omega<\omega _{c}$, $n_{M}$ is real and negative since the square root in (\ref{n-R-M-indices-1}) is real. This is the same behavior of the index {$-n_{-}$}. See the black line in Fig. (\ref{nMeta}).  
	
	\item For $\omega\rightarrow\omega _{c}$, $n_{M}\rightarrow-\infty$,
	and there occurs a resonance at the cyclotron frequency.
	
	\item For $\omega_{c}<\omega<\omega_{r}$, there appears an absorption zone for metamaterial, $\mathrm{Re}[n_{M}]<0$ and $\mathrm{Im}[n_{M}]\ne0$, while the index {$-n_{-}$} is purely imaginary, $\mathrm{Re}[n_{M}]=0$ and $\mathrm{Im}[n_{M}]\ne0$, as shown in Fig.~\ref{nMeta}. The frequency $\omega_{r}$ is the root of $R_{-}$. 
	
	\item For $\omega>\omega _{r}$, the quantity $n_{M}$ is always negative, corresponding to a negative propagation zone, with $n_{M} \rightarrow -1$ in the high-frequency limit.
	
\end{enumerate}

{
\subsection{\label{section-helicons}Low-frequency modes}
Considering the low-frequency regime,
\begin{align}
\omega\ll \omega_{p}, \quad \omega_{c}\ll\omega_{p}, \quad  \omega\ll \omega_{c}, \label{helicon-frequency-regime}
\end{align}
{the magnetized plasma RCP refractive index (\ref{nusual2}) provides a helicon mode,} described by 
\begin{align}
n_{-} &= \omega_{p} \sqrt{\frac{1}{\omega \omega_{c}}} . \label{helicons-12}
\end{align}}
{Helicons are RCP modes that propagate at very low frequencies and along the magnetic field axis. See Ref. \cite{Bittencourt} {(Chapter 9)}, and Ref. \cite{chapter-8} {(Chapter 8)} for basic details.} 
{
For the {electromagnetic} modes obtained in \eqref{n-R-M-indices-1} and \eqref{n-L-E-indices-1}, the corresponding helicons indices are 
\begin{align}
\bar{n}_{R,E}&=\frac{\omega_{p}^{2}}{\omega_{c} V_{0}}, \label{helicons-15} \\
\bar{n}_{L,M} &=-\frac{\omega_{p}^{2}}{\omega_{c} V_{0}} \pm \frac{V_{0}}{\omega} , \label{helicons-16}
\end{align}
where we have used the ``bar'' notation to indicate the helicons quantities.} {In this chiral context, we observe the existence of both RCP and LCP helicons, that both propagate in the {low-energy} regime, due to the presence of the timelike component, $V_{0}$. The helicon modes given by $\bar{n}_{R,E}$ have constant and {nondispersive} indices, which depend on the inverse of $V_{0}$. On the other hand, {negative-refraction} dispersive helicons are associated with $\bar{n}_{M}$, and also with $\bar{n}_{L}$ for $\omega<\omega_{c} V_{0}^2/\omega_{p}^{2}$.  Note that the term $V_{0}/\omega$ appears as a channel of distinction between the helicons associated with $\bar{n}_{R,E}$ and $\bar{n}_{L,M}$, as expected. }

{
The cold plasma usual helicon mode is recovered if one also considers, besides the relations  (\ref{helicon-frequency-regime}), the condition $V_{0} \ll \omega$. In this case, the expansion of $n_{R, E}$ of \eqref{n-R-M-indices-1} and $n_{L, M}$ of \eqref{n-L-E-indices-1} yields
 \begin{align}
\tilde{n}_{R,M} &=  \pm \omega_{p} \sqrt{\frac{1}{\omega \omega_{c}}}, \label{helicons-extra-small-LV-1}
\end{align}
 {which is} the same result {as} the usual case {[see \eqref{helicons-12}]}, {and $\tilde{n}_{L, E}$ being purely imaginary}. Here, the ``tilde'' notation indicates helicons in the specific case where $\omega_{c} \ll \omega_{p}$, $\omega \ll \omega_{c}$, and $V_{0} \ll \omega$. }

\subsection{Dispersion relations behavior}

The wave dispersion associated with each refractive index is usually visualized in plots $\omega \times k$. In the following, we work with dimensionless plots,  $(\omega/\omega_{c}) \times (k/\omega_{c})$.

\begin{figure}[h]
	\centering
	\includegraphics[scale=0.5]{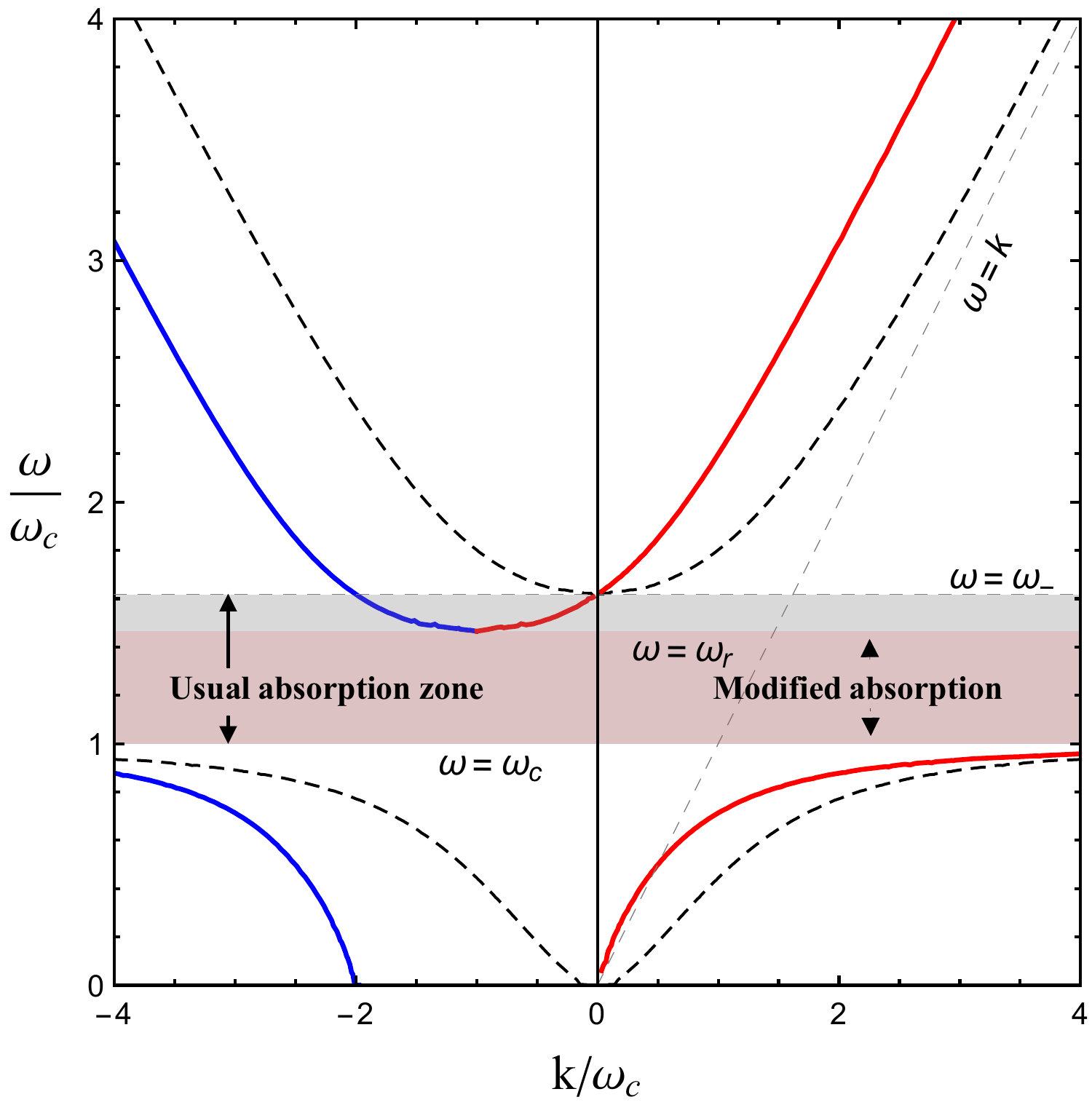}  \caption{Plot of the dispersion relations related to refractive indices $n_{R}$ (solid red line) and $n_{M}$ (solid blue line). The dashed black line corresponds to the indices of the usual case ($\pm n_{-}$). The highlighted area in red (gray) indicates the absorption zone for $n_{R,M}$ ($\pm n_{-}$). Here, we have used $\omega_{c}=\omega_{p}$ and $V_{0}=2\omega_{p}$, with $\omega_{c}=1$~$\mathrm{rad}$~$s^{-1}$.}
	\label{oxk_nr}%
\end{figure}

The dispersion relations associated with $n_{R}$ and $n_{M}$ are depicted in Fig.~\ref{oxk_nr} for $\omega_{c}=\omega_{p}$. The propagation occurs for $0<\omega<\omega_{c}$ and $\omega>\omega_{-}$, while absorption takes place in $\omega_{c}<\omega<\omega_{r}$. The range $\omega_{r}<\omega<\omega_{-}$ corresponds to negative refraction propagation zone ($k<0$) for $n_{R}$. The refractive index $n_{M}$ is negative for $k<0$ and $\omega>0$.

\begin{figure}[h]
	\centering
	\includegraphics[scale=0.5]{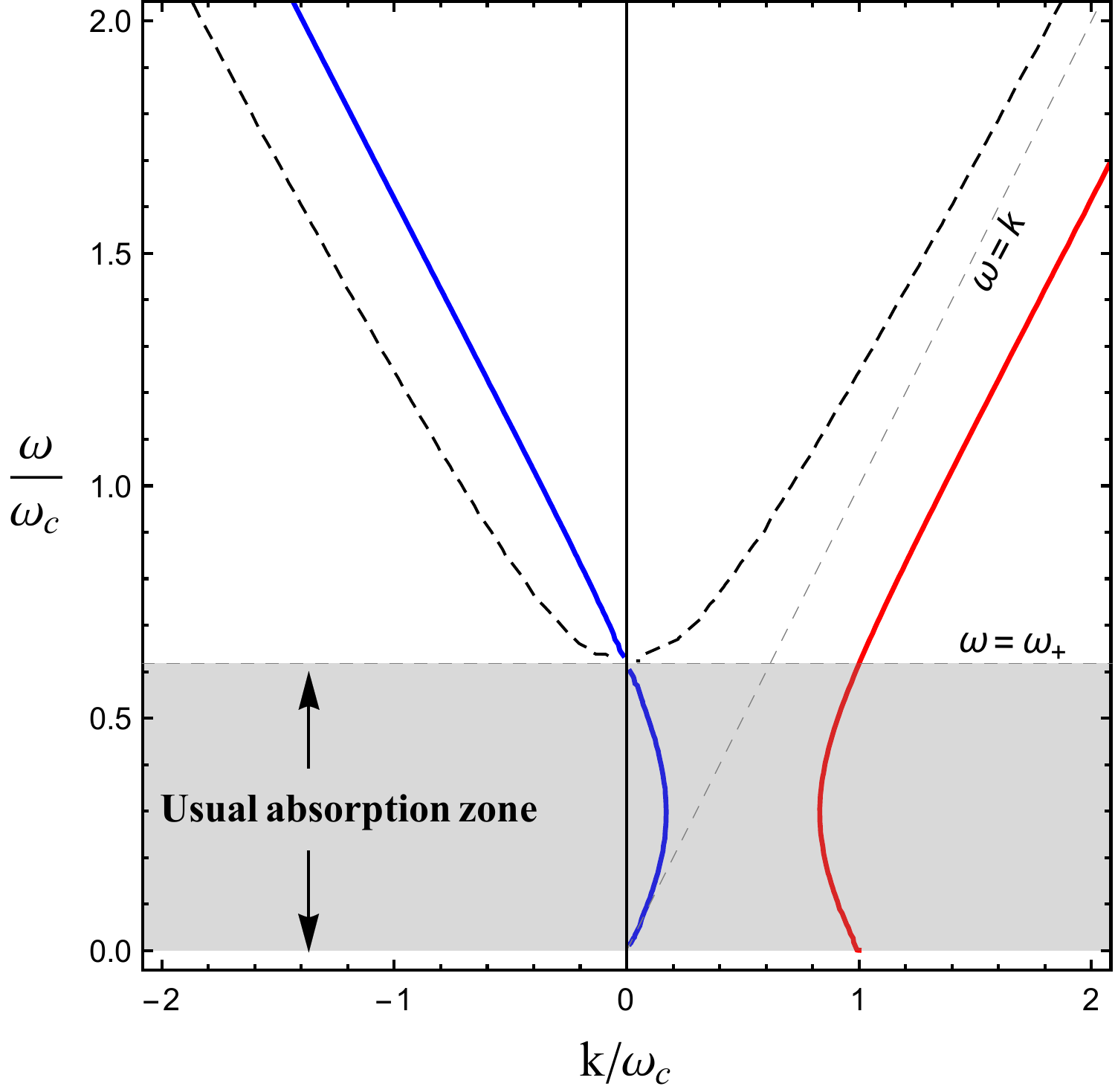}  \caption{Plot of the dispersion relations related to refractive indices $n_{L}$ (solid red line) and $n_{E}$ (solid blue line). The dashed line corresponds to the indices $\pm n_{+}$ of the usual case. The highlighted gray area indicates the absorption zone for $\pm n_{+}$, where now also occurs propagation. Here, we have used $\omega_{c}=\omega_{p}$ and $V_{0}=\omega_{p}$, with $\omega_{c}=1$~$\mathrm{rad}$~$s^{-1}$.}
	\label{oxk_nl}%
\end{figure}

Figure \ref{oxk_nl} depicts the dispersion relations related to $n_{L}$ and $n_{E}$. The wave associated with $n_{L}$ propagates for all frequencies. For $n_{E}$, the conventional propagation zone occurs in $0<\omega<\omega_{+}$. For $\omega>\omega_{+}$, there occurs a propagation zone with negative refraction.  For the standard indices, $\pm n_{+}$, the absorption zone is $0<\omega<\omega_{+}$.

\begin{figure}[h]
	\centering
	\includegraphics[scale=0.5]{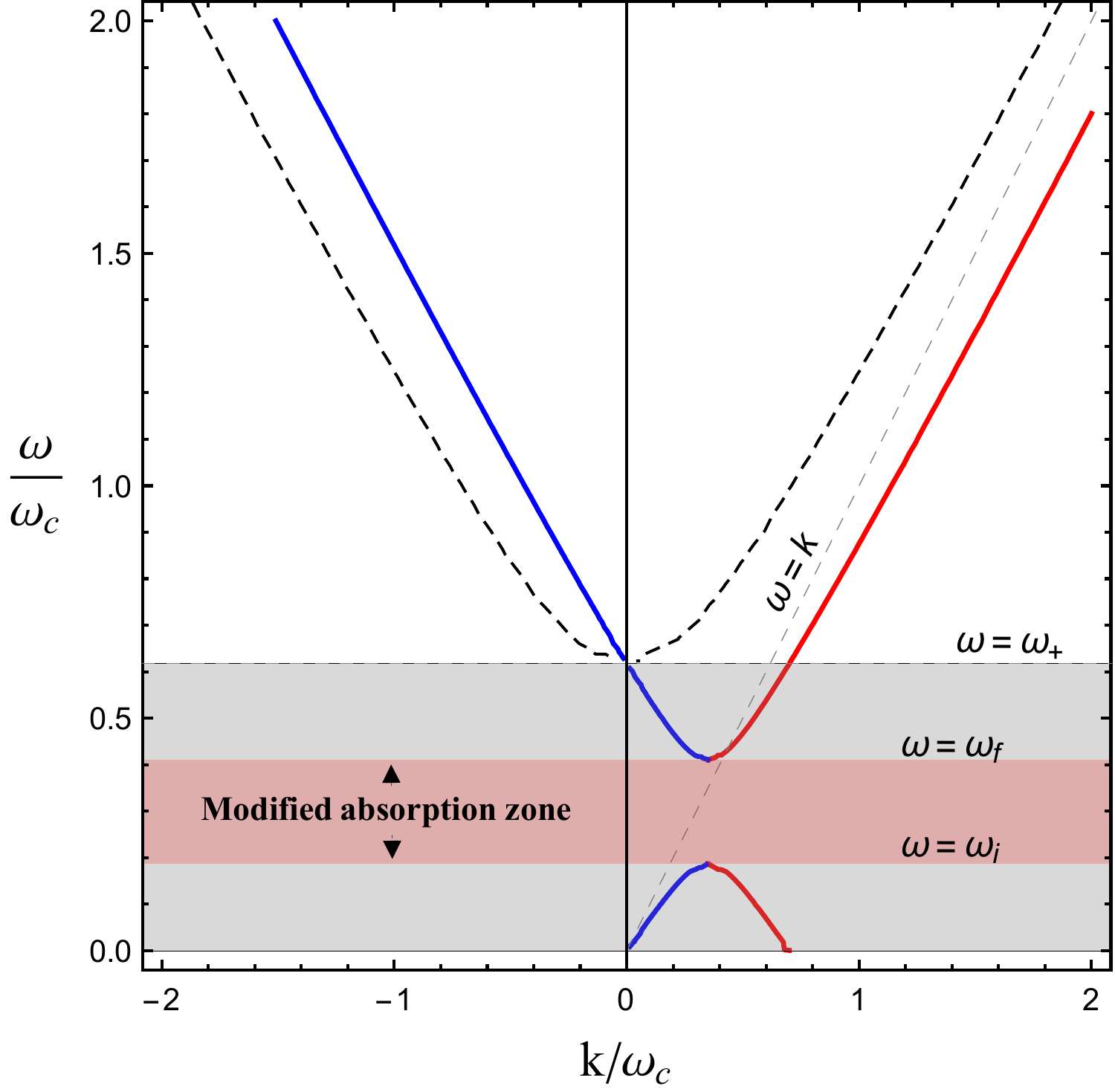}  \caption{Plot of the dispersion relations related to refractive indices $n_{L}$ (solid red line) and $n_{E}$ (solid blue line). The dashed line corresponds to the usual case with indices $\pm n_{+}$. The highlighted areas in red (gray) indicate the absorption zone for $n_{L,E}$ ($\pm n_{+}$). Here, we have used $\omega_{c}=\omega_{p}$ and $V_{0}=0.7\omega_{c}$, with $\omega_{c}=1$~$\mathrm{rad}$~$s^{-1}$.}
	\label{oxk_nl2}%
\end{figure}

Furthermore, Fig.~\ref{oxk_nl2} shows the dispersion relations for $n_L$ and $n_{E}$ in the case there is a modified absorption zone for $\omega_{i}<\omega<\omega_{f}$, while the free propagation occurs for $0<\omega<\omega_{i}$ and $\omega>\omega_{f}$. The frequencies $\omega_{i}$, $\omega_{f}$, and $\omega_{r}$ define the limits for unusual propagation zones. As already discussed, these frequencies are obtained from the radicands (\ref{radnR}) and (\ref{rad}).

{
\section{WAVE PROPAGATION ORTHOGONAL TO THE MAGNETIC FIELD\label{sec5}} 
}

{For investigating wave propagation orthogonally to the magnetic field direction, we rewrite \eqref{timelike2} for $\mathbf{n}=\left(n_{x},n_{y},0\right)$, that is,   }

{
	\begin{equation}
	\begin{bmatrix}
	n^{2}-n_{x}^{2}-S & iD-n_{x}n_{y} &
	-i\left(V_{0}/\omega\right)n_{y}\\
	-iD-n_{x}n_{y} & n^{2}-n_{y}^{2}-S & +i\left(V_{0}/\omega\right)n_{x}\\
	+i\left(V_{0}/\omega\right)n_{y} & -i\left(V_{0}/\omega\right)n_{x} & n^{2} -P
	\end{bmatrix}
	\begin{bmatrix}
	\delta E_{x}\\
	\delta E_{y}\\
	\delta E_{z}%
	\end{bmatrix}
	=0. \label{timelike3}
	\end{equation}
	}
	
{Using the parametrization  $\mathbf{n}=n\left(\cos\phi,\sin\phi,0\right)$, it becomes}

{\begin{widetext}	
		\begin{equation}
		\begin{bmatrix}
		n^{2}-n^{2}\cos^2\phi-S & iD-n^{2}\sin\phi\cos\phi &
		-i\left(V_{0}/\omega\right)n\sin\phi\\
		-iD-n^{2}\sin\phi\cos\phi & n^{2}-n^{2}\sin^2\phi-S & +i\left(V_{0}/\omega\right)n\cos\phi\\
		+i\left(V_{0}/\omega\right)n\sin\phi & -i\left(V_{0}/\omega\right)n\cos\phi & n^{2} -P
		\end{bmatrix}
		\begin{bmatrix}
		\delta E_{x}\\
		\delta E_{y}\\
		\delta E_{z}%
		\end{bmatrix}
		=0, \label{timelike4}
		\end{equation}
	\end{widetext}
whose null determinant yields the angle-independent dispersion relation,
	\begin{equation}
	\frac{n^2 S V_{0}^2}{\omega ^2}-\left(n^2-P\right) \left(D ^2+S \left(n^2-S\right)\right)=0,
	\end{equation}
providing two refractive indices given by
	\begin{equation}
	n_{O\pm}^2=\frac{P+S}{2}-\frac{D ^2}{2S}+\frac{V_{0}^2}{2\omega ^2} \pm\frac{\Gamma}{2S}, \label{indicesortogonal}
	\end{equation}
where
	\begin{equation}
	\Gamma=\sqrt{\left(D ^2-P S-S^2-\frac{S V_{0}^2}{\omega ^2}\right)^2-4 S P\left( S^2-D ^2 \right)}.
	\end{equation}
	From \eqref{timelike4}, the indices in \eqref{indicesortogonal} are related to the following propagating modes:
	\begin{equation}
	 \mathbf{{E}}=C \begin{bmatrix}
	\zeta \\ 
	1 \\
	i\frac{SV_{0}n_{O\pm}}{\gamma\omega\left(n_{O\pm}^2-P\right)}
	\end{bmatrix},\label{autovetor}
	\end{equation} 
with 
	\begin{equation}
C=\frac{\omega\left(n_{O\pm}^2-P\right)\left|\gamma\right|}{\sqrt{2\omega^2\left(n_{O\pm}^2-P\right)^2\left|\gamma\right|^2+\left(SV_{0}n_{O\pm}\right)^2}},
	\end{equation}
	\begin{equation}
	\zeta=-\frac{S\sin\phi-iD\cos\phi}{\gamma},
	\end{equation}
	\begin{equation}
	\gamma=S\cos\phi+iD\sin\phi.
	\end{equation} 
	The refractive index $n_{O+}$ has a cutoff {at the plasma frequency} $\omega_{p}$, {while $n_{O-}$} has two cutoff frequencies, $\omega_{\pm}$, given in \eqref{r1}. }

\subsection{About the index $n_{O+}$ \label{secNort1}} 
\begin{figure}[h]		
	\centering
	\includegraphics[scale=0.65]{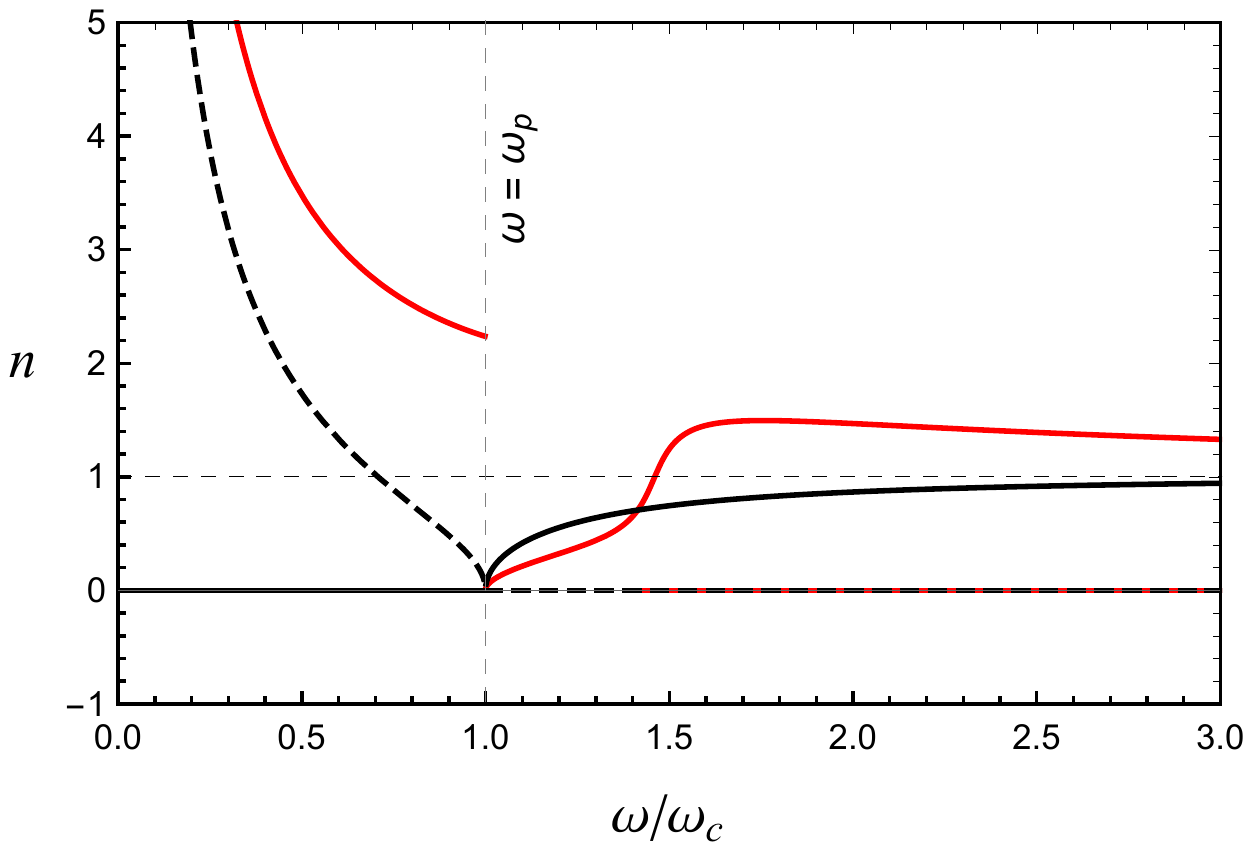}
	\caption{Red line: plot of the index $n_{O+}$. Black line: plot of the index $n_{T}$. Dashed (solid) lines represent the imaginary (real) pieces of $n_{O+}$ and $n_{T}$. Here, we have used: $\omega_{c}=\omega_{p}$, $V_{0}=2\omega_{p}$, and $\omega_{c}=1$~$\mathrm{rad}$~$s^{-1}$.}
	\label{nort1}
\end{figure}

{The refractive index $n_{O+}$ has to be compared to the index $n_{T}$, given in \eqref{nusualort1}, associated with the usual transversal mode, since in the limit $V_{0}\rightarrow 0$ {it recovers the refractive} index $n_{T}$.  We present below some aspects of $n_{O+}$: }
	\begin{enumerate}
		[label=(\roman*)]
		
      \item 	{For $0<\omega<\omega _{p}$, there is a propagation zone in which $n_{O+}$ is real. This behavior is markedly different from the usual case, for which there corresponds to an absorption zone (in this range). See the dashed black line in Fig.~\ref{nort1}.  }

		\item {For $\omega\rightarrow\omega _{p}$, $n_{O+}$ has an unusual discontinuity, as shown in Fig. \ref{nort1} (see the red curve). }
		
		\item {For $\omega>\omega _{p}$, the index} $n_{O+}$ is always real, corresponding to a propagation zone.
\end{enumerate}

 \subsection{About the index $n_{O-}$ \label{secNort2}} 
\begin{figure}[h]		
 	\centering
 	\includegraphics[scale=0.65]{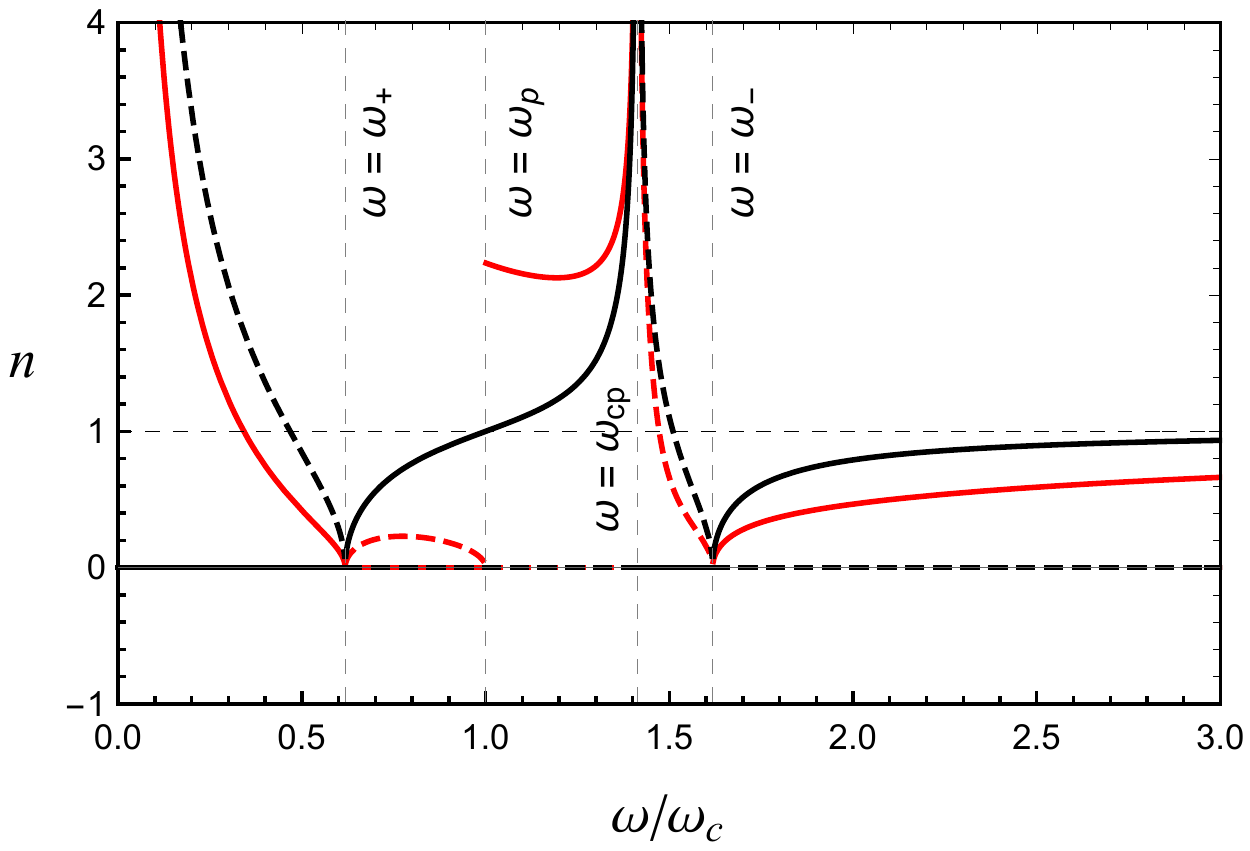}
 	\caption{Red line: plot of the index $n_{O-}$. Black line: plot of the index $n_{O}$. Dashed (solid) lines represent the imaginary (real) pieces of $n_{O-}$ and $n_{O-}$. Here, we have used $\omega_{c}=\omega_{p}$, $V_{0}=2\omega_{p}$, and $\omega_{c}=1$~$\mathrm{rad}$~$s^{-1}$.}
 	\label{nort2}
 \end{figure}

 {The refractive index $n_{O-}$, given in \eqref{indicesortogonal}, is a modification  {of the index (\ref{nusualort2})}, associated with the usual {extraordinary} mode. The index $n_{O-}$ shares the same usual resonance frequency, \begin{equation}
 	\omega_{cp}=\sqrt{\omega_{c}^2+\omega_{p}^2.}
 	\end{equation}
 	We point out:}
 \begin{enumerate}
 	[label=(\roman*)]
 	
 	\item {For $0<\omega<\omega _{+}$, $n_{O-}$ is real and positive, corresponding to a propagation zone. It contrasts with the usual case, where $n_{O}$ is imaginary in this range (absorption zone). See the black line in Fig. \ref{nort2}.  } 
 	
 	\item {For $\omega _{+}<\omega<\omega_{p}$, there occurs an absorption zone, where $\mathrm{Re}[n_{O-}]=0$ and $\mathrm{Im}[n_{O-}]\ne0$. In the standard case, there is a propagation zone in this range.}
 	
 		\item {For $\omega\rightarrow\omega _{p}$, $n_{O-}$ has a discontinuity, as shown in Fig. \ref{nort2}. For $\omega _{p}<\omega<\omega_{cp}$, the index $n_{O-}$ is real and there appears a propagation zone. The same positivity occurs in the usual case. }
 		
 		 \item {For $\omega\rightarrow\omega _{cp}$, the index $n_{O-}\rightarrow+\infty$ and there occurs a resonance. For $\omega _{cp}<\omega<\omega_{-}$,  $\mathrm{Re}[n_{O-}]=0$ and $\mathrm{Im}[n_{O-}]\ne0$, and one has an absorption zone, the same behavior of the usual case in this range.}
 	
 	\item {For $\omega>\omega _{-}$, the quantity $n_{O-}$ is always positive, corresponding to a propagation zone.}
 	
 \end{enumerate}
 
\section{Birefringence, rotatory power and dichroism \label{birefringence}}

The phase velocity in terms of the refractive index $n$ is defined (in natural units) as $v_{phase}=1/n$. Hence, the corresponding phase velocities, $v_{R}=1/\left(n_{R}\right)$, $v_{L}=1/\left(n_{L}\right)$,
	$v_{E}=1/\left(n_{E}\right)$, $v_{M}=1/\left(n_{M}\right)$, can be defined
with the indices $n_{R}$, $n_{L}$, $n_{E}$, $n_{M}$ of Eqs. (\ref{n-R-M-indices-1}) and (\ref{n-L-E-indices-1}). Accordingly with the previous analysis of the refractive indices, in general, the RCP and LCP modes propagate at different phase velocities for each frequency value, generating circular birefringence in the propagation band, expressed in terms of the rotatory power (\ref{powerROT}). On the other hand, in the absorption zones, there occurs dichroism, measured in terms of the coefficient of \eqref{dicro}.

\subsection{Rotatory power \label{secRP}}

In order to write the rotatory power, we need to consider the refractive indices $n_{L}$, $n_{E}$, associated with the LCP wave, and the indices $n_{R}$, $n_{M}$, associated to the RCP wave. It allows, in principle, to determine four distinct RPs at the propagation zones, some of which we examine in this section.

\begin{figure}[h]
	\centering
	\includegraphics[scale=0.7]{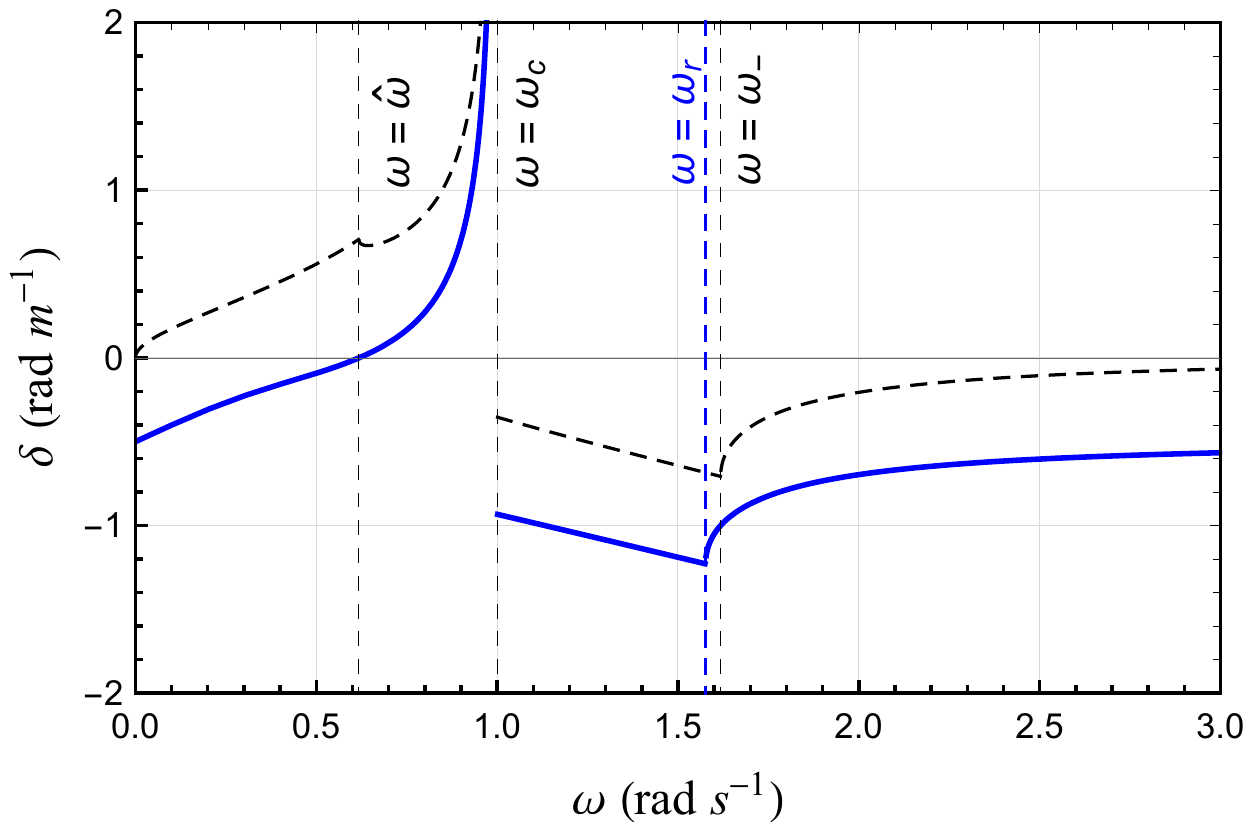}
	\caption{The solid blue line represents the rotatory power (\ref{rptl}) defined by the refractive index $n_{L}$ and $n_{R}$, for the condition (\ref{cond}). The dashed black line corresponds to the usual rotatory power (\ref{rpcasousual1}). Here, we have used $\omega_{c}=\omega_{p}$, $V_{0}=\omega_{p}$, and $\omega_{c}=1$~$\mathrm{rad}$~$s^{-1}$.} 
	\label{rptl1}
	
\end{figure}

We start by writing the rotation power defined in terms of real pieces of the refractive indices $n_{L}$ and $n_{R}$, 
\begin{equation}
\delta_{LR}=-\frac{\omega}{2}\left(\mathrm{Re}[n_{L}]-\mathrm{Re}[n_{R}]\right), 
\end{equation}
or explicitly,
\begin{equation}
\delta_{LR}=-\frac{\omega}{2}\mathrm{Re}\left[V_{0}/\omega+\sqrt{R_{+}}-\sqrt{R_{-}}\right], \label{rptl}
\end{equation}
where $R_{+}$ and $R_{-}$ are given in Eqs. (\ref{radnR}) and (\ref{rad}). We find a positive frequency,
\begin{equation}
\hat{\omega}=\sqrt{\omega_{c}^2+\omega_{p}^2/2-\frac{\omega_{p}^2 \sqrt{4 \omega_{c}^2+V_{0}^{2}}}{2V_{0}}},
\end{equation}
where the RP (\ref{rptl}) undergoes a sign reversal.  In Fig. \ref{rptl1}, we illustrate the behavior of RP for the condition (\ref{cond}). For the interval $0<\omega<\hat{\omega}$, the RP is negative, and for $\hat{\omega}<\omega<\omega_{c}$, it is positive. The RP reversion that occurs at $\omega=\hat{\omega}$ is not usual in cold plasmas theory. However, it is reported in graphene systems \cite{Poumirol}, rotating plasmas \cite{Gueroult}, and bi-isotropic dielectrics supporting chiral magnetic current \cite{PedroPRB}. For $\omega>\omega_{c}$, the RP is always negative. Nevertheless, it is necessary to pay attention to the interval $\omega_{c}<\omega<\omega_{r}$, where the refractive index $n_{R}$ has an imaginary piece and the RCP wave is absorbed. At $\omega=\omega_{r}$,  the real piece of $n_{R}$ undergoes a sharp change (see Fig.~\ref{nRfig}), which also appears in the RP profile of Fig.~\ref{rptl1}.

We can safely claim that both modes associated with the $n_{L}$ and $n_{R}$ propagate for $\omega>\omega_{-}$, range in which the RP magnitude decreases monotonically with $\omega$, approaching to its asymptotic value, $-V_{0}/2$ (see Fig.~\ref{rptl1}). Assuming the limit where $\omega>>\left( \omega_{p},\omega_{c}\right) $, we can write

\begin{align}
n_{L,R} &\approx 1 \pm \frac{V  _{0}}{2\omega}+\frac{V _{0}^{2}}{8\omega^{2}}-\frac{\omega_{p}^{2}}{2\omega\left(
	\omega\pm\omega_{c}\right)  }, 
\end{align}
so that the rotatory power is
\begin{equation}
\delta_{LR}\approx -\frac{ V  _{0}}{2}-\frac{\omega_{p}^{2}%
	\omega_{c}}{ 2\omega^{2}}.\label{faradayTIMELIKE}%
\end{equation}

Note that taking the limit $V_{0}\rightarrow0$, the usual Faraday effect RP (\ref{rotFaraday}) is recovered for the high-frequency regime. It is also interesting to point out that the Faraday effect disappears for a null magnetic field, $\omega_{c}=0$. However, the birefringence still remains, due to the presence of the chiral term, which yields the following RP:
	\begin{equation}
	\delta\approx -V_{0}/2. \label{faradayTIMELIKE1}
	\end{equation}
	
For the condition (\ref{cond2}), the RP (\ref{rptl}) also exhibits a sign reversal and a very similar profile to the one of Fig.~\ref{rptl1}, in such a way that it will not be depicted here.

\begin{figure}[h]
	
	\centering
	\includegraphics[scale=0.7]{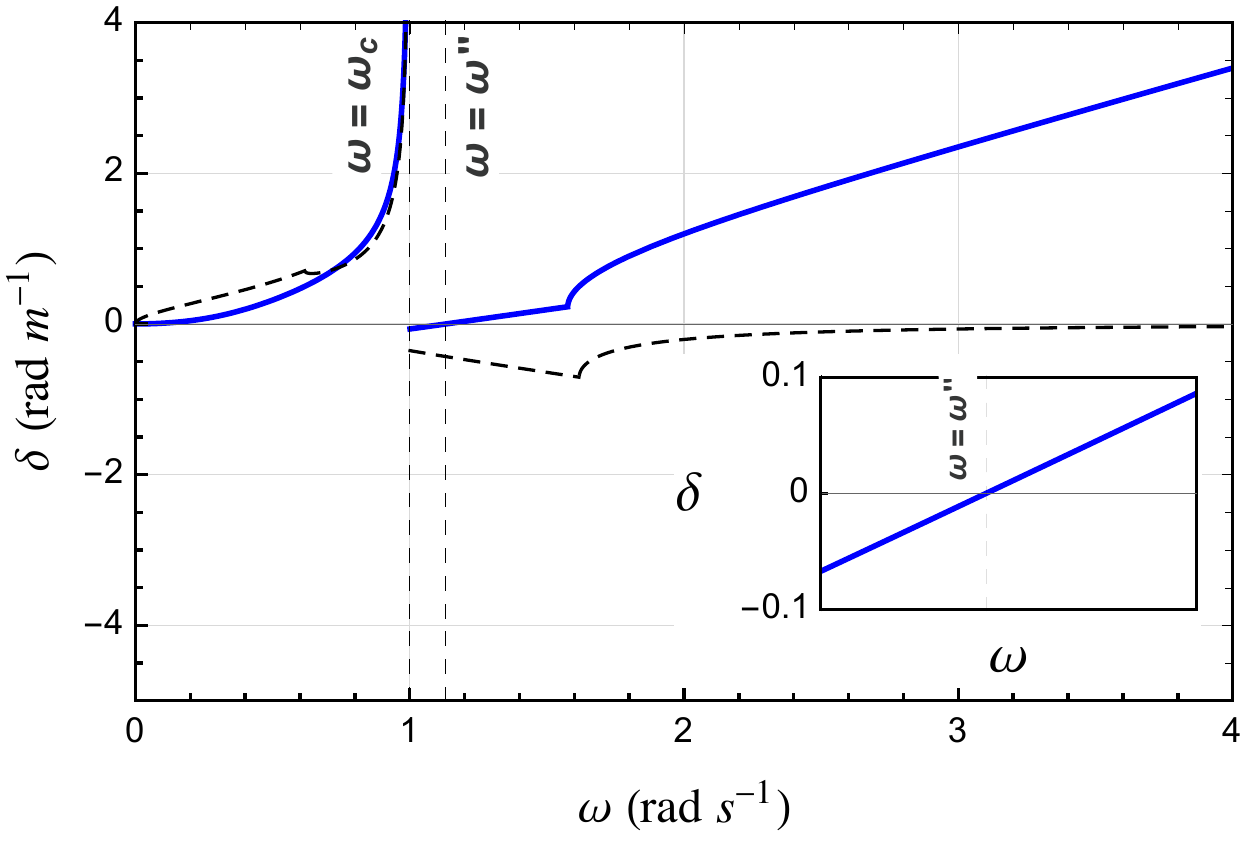}
	\caption{Solid blue lines: plot of the rotatory power (\ref{rpeq2}) associated to the refractive indices $n_{E}$ and $n_{R}$ for the condition (\ref{cond}). The dashed line represents the usual rotatory power (\ref{rpcasousual1}). Here, we have used $\omega_{c}=\omega_{p}$, $V_{0}=\omega_{p}$, and $\omega_{c}=1$~$\mathrm{rad}$~$s^{-1}$. The inset plot highlights the behavior of $\delta$ around $\omega=\omega''$.} 
	\label{rptl1NE}
	
\end{figure}

\begin{figure}[t]
	
	\centering
	\includegraphics[scale=0.7]{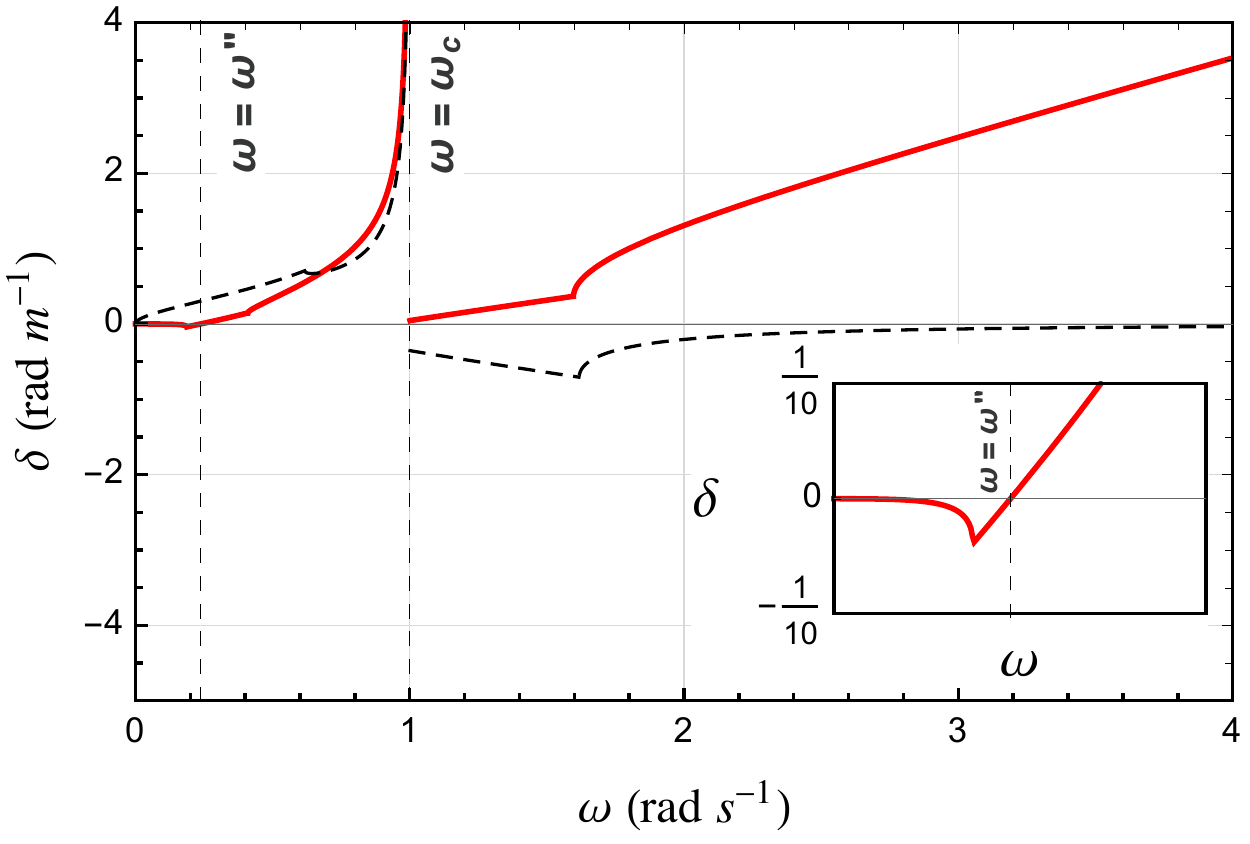}
	\caption{Solid red lines: rotatory power (\ref{rpeq2}) associated to the refractive indices $n_{E}$ and $n_{R}$ for the condition (\ref{cond2}). The dashed line represents the usual rotatory power (\ref{rpcasousual1}). Here, we have used $\omega_{c}=\omega_{p}$, $V_{0}=0.7\omega_{p}$, and $\omega_{c}=1$~$\mathrm{rad}$~$s^{-1}$. The inset plot highlights the behavior of $\delta$ around $\omega=\omega''$.} 
	\label{rptl1NE2}
\end{figure}

Considering now the refractive indices $n_{E}$ and $n_{R}$, the rotatory power is
\begin{equation}
\delta_{ER}=-\frac{\omega}{2}\left(\mathrm{Re}[n_{E}]-\mathrm{Re}[n_{R}]\right), 
\end{equation}
or,
\begin{equation}
\delta_{ER}=-\frac{\omega}{2}\mathrm{Re}\left[V_{0}/\omega-\sqrt{R_{+}}-\sqrt{R_{-}}\right].\label{rpeq2} 
\end{equation}
Recalling that the LCP wave associated with $n_{E}$ has a conventional free propagation for $\omega<\omega_{+}$ and propagation with negative refractive index ($n_{E}<0$) for $\omega>\omega_{+}$ (with $\omega_{+}<\omega_{c}$), the RP magnitude is enhanced in the latter zone. This behavior is depicted in Fig.~\ref{rptl1NE}, which shows the RP (\ref{rpeq2}) for $n_{E}$ given by the condition (\ref{cond}), $R_{+}>0$. The RP is positive for $\omega<\omega_{c}$ and negative for $\omega_{c}<\omega<\omega^{\prime\prime}$, becoming positive again for $\omega >\omega''$, where $\omega^{\prime\prime}$ is the reversal frequency. For $n_{E}$ given by the condition condition (\ref{cond2}), the RP is depicted in Fig. \ref{rptl1NE2}, revealing a small reversion at $\omega^{\prime\prime}<\omega_{c}$.  Note that the increasing RP with $\omega$, depicted in Figs. \ref{rptl1NE} and \ref{rptl1NE2}, is due to the negative behavior of the index $n_{E}$ for $\omega>\omega_{+}$, {that is, enhancement associated with the negative refraction.}

In the asymptotic limit, where $\omega>>\left( \omega_{p},\omega_{c}\right) $, the RP (\ref{rpeq2}) goes as
	\begin{equation}
		\delta_{ER}\approx \omega-\frac{ V _{0}}{2},\label{faradayTIMELIKE2}
	\end{equation}
	presentig a predominant linear behavior in $\omega$, as it appears  in Figs. \ref{rptl1NE} and \ref{rptl1NE2}. It is also worth mentioning that the limit $V_{0}\rightarrow0$, implying $\delta\approx\omega$, does not stand for a valid result for usual magnetized plasma, since the RP (\ref{rpeq2}) is not defined for achiral cold plasmas.

\subsection{Dichroism coefficients \label{secDC} }
\begin{figure}[h]	
	\centering
	\includegraphics[scale=0.65]{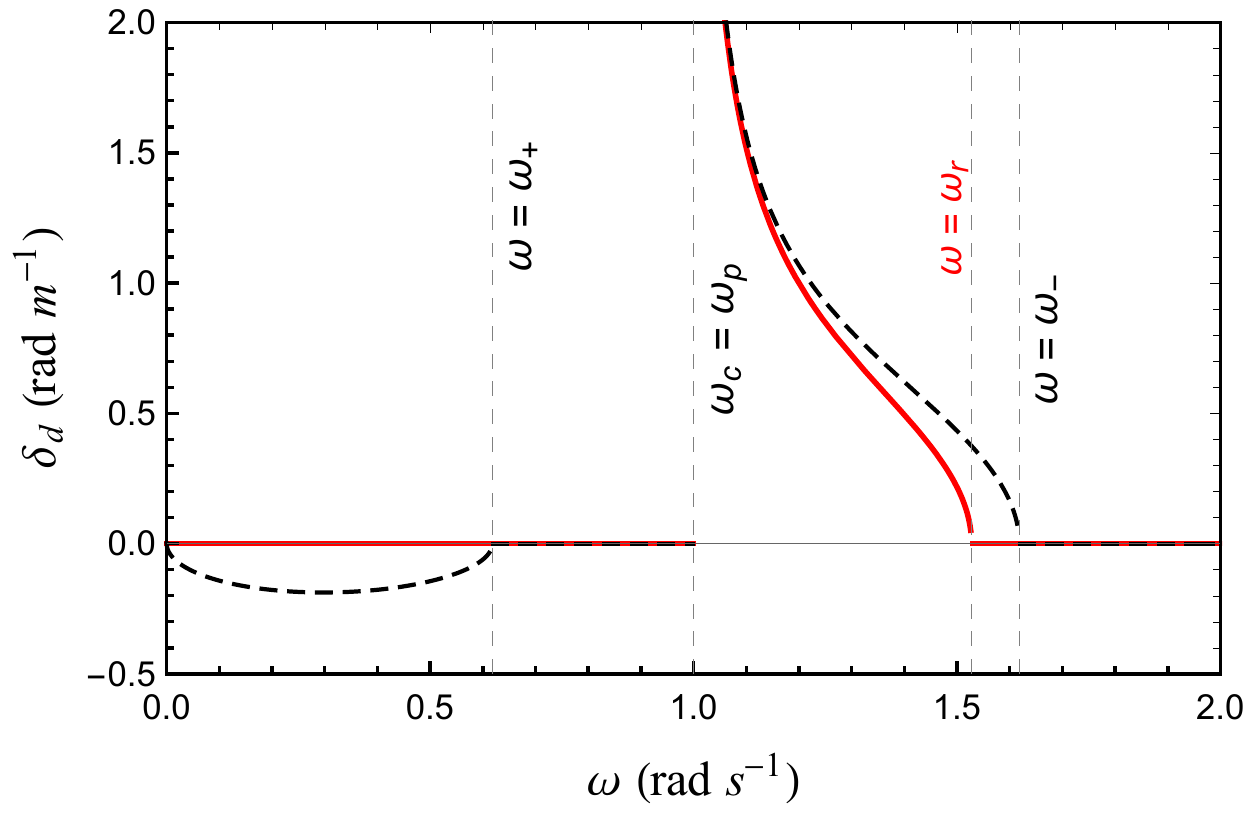}
	\caption{Plot of the dichroism coefficient (\ref{dicro_TL2.2})(red solid lines) associated to the refractive indices $n_{L}$ and $n_{R}$, under the condition (\ref{cond}). The black dashed line represents the usual dichroism coefficient (\ref{dicrousualcase}). Here $\omega_{c}=\omega_{p}$, $V_{0}=\left(3/2\right)\omega_{c}$, and $\omega_{c}=1$~$\mathrm{rad}$~$s^{-1}$.}
	\label{dicro_tlplot1}
\end{figure}

As well known, absorption depends on the magnitude of the imaginary parts of the refractive indices. When one mode is more absorbed than the other, there occurs dichroism. Considering the refractive indices $n_{L}$ and $n_{R}$, the circular dichroism coefficient is
\begin{equation}
\delta_{dLR} =-\frac{\omega }{2}\left( \mathrm{Im}[n_{L}]-\mathrm{Im}[n_{R}]\right).  \label{dicro_eqtl}
\end{equation}
Considering the condition (\ref{cond}), only $n_{R}$ has imaginary part (localized in the interval $\omega_{c}<\omega<\omega_{-}$), while $n_{L}$ is real for $\omega>0$. In this case, the dichroism coefficient is given by
\begin{equation}
\delta_{dLR}=\begin{cases}
\text{$0,$} &\quad\text{for $0<\omega<\omega_{c},$}\\
\text{$\sqrt{R_{-}},$} &\quad\text{for $\omega_{c}<\omega<\omega_{r},$}\\
\text{$0,$} &\quad\text{for $\omega>\omega_{r},$}
\end{cases}\label{dicro_TL2.2}
\end{equation}
being {non-null} only in the range $\omega_{c}<\omega<\omega_{-}$, as properly shown in Fig~\ref{dicro_tlplot1}. 

\begin{figure}[h]	
	\centering
	\includegraphics[scale=0.65]{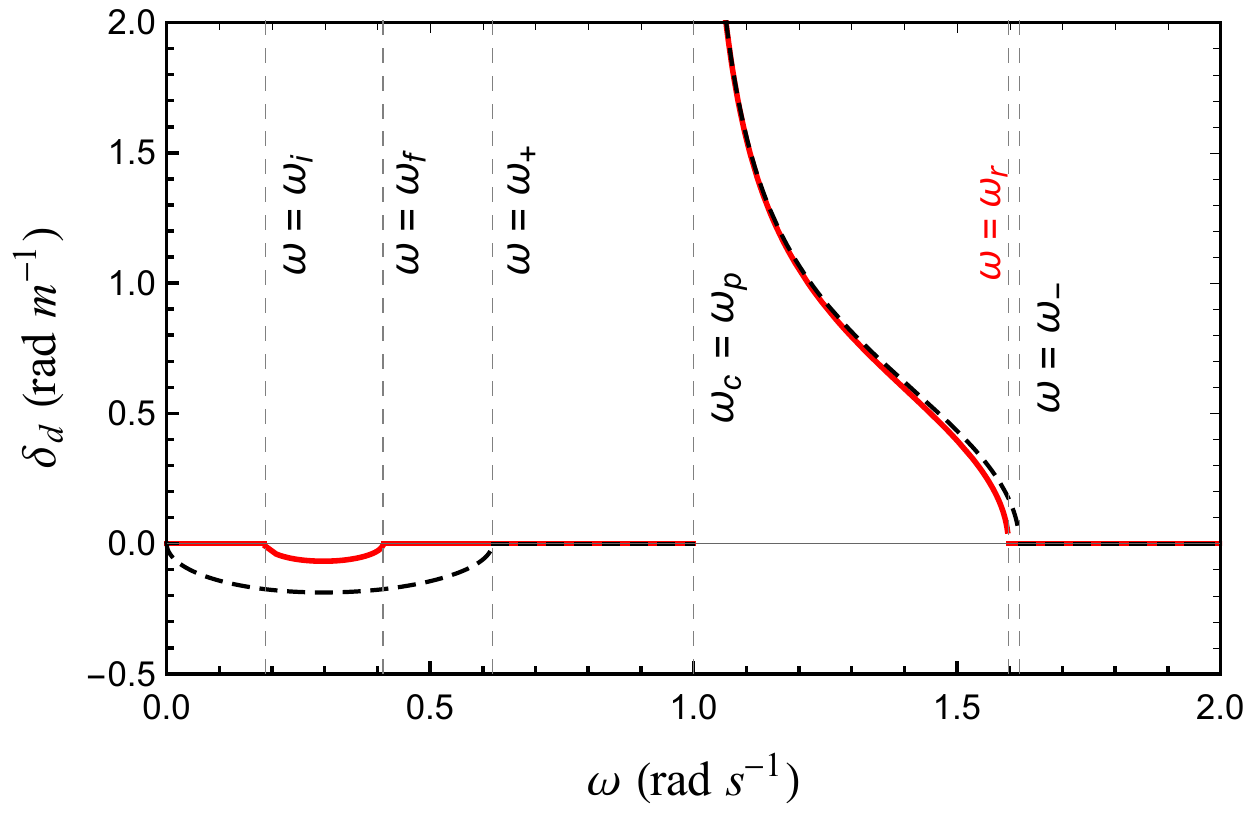}
	\caption{Plot of the dichroism coefficient (\ref{dicro_TL2.3})(solid red lines) associated to the refractive indices $n_{L}$ and $n_{R}$, under the condition (\ref{cond2}). The dashed line represents the usual dichroism coefficient (\ref{dicrousualcase}). Here, we have set $\omega_{c}=\omega_{p}$, $V_{0}=0.7\omega_{p}$, and $\omega_{c}=1$~$\mathrm{rad}$~$s^{-1}$.}
	\label{dicrotlplot2}
\end{figure}

Considering the condition (\ref{cond2}), both $n_{R}$ and $n_{L}$ have non-null imaginary parts in the intervals $\omega_{c}<\omega<\omega_{r}$ and $\omega_{i}<\omega<\omega_{f}$, respectively. The dichroism coefficient is null for $0<\omega< \omega_{i}$, $\omega_{f} < \omega < \omega_{c}$, and $\omega >\omega_{r}$, being non-null only for
\begin{equation}
\delta_{dLR}=\begin{cases}
\text{$-\frac{\omega}{2} \sqrt{R_{+}}$}, &\quad\text{for $\omega_{i}<\omega<\omega_{f}$},\\
\text{$+\frac{\omega}{2} \sqrt{R_{-}}$}, &\quad\text{for $\omega_{c}<\omega<\omega_{r}$},\\
\end{cases}     \label{dicro_TL2.3}
\end{equation}
whose general behavior is exhibited in Fig.~\ref{dicrotlplot2}. 

For the refractive indices $n_{E}$ and $n_{R}$, the circular dichroism coefficient is
\begin{equation}
\delta_{dER} =-\frac{\omega }{2}\left( \mathrm{Im}[n_{E}]-\mathrm{Im}[n_{R}]\right).  \label{dicro_eqtl2}
\end{equation}
If we consider $n_{E}$ under the condition (\ref{cond}), the same behavior of Fig.~\ref{dicro_tlplot1} is obtained, since $n_{E}$ is always real, not contributing to the dichroism.
On the other hand, regarding the condition (\ref{cond2}), both $n_{R}$ e $n_{E}$ have {nonzero} imaginary parts in the intervals $\omega_{c}<\omega<\omega_{r}$ and $\omega_{i}<\omega<\omega_{f}$, respectively. In this case, we have
\begin{equation}
\delta_{dER}=\begin{cases}
\text{$0$}, &\quad\text{for $0<\omega<\omega_{i}$},\\
\text{$+\frac{\omega}{2} \sqrt{R_{+}}$}, &\quad\text{for $\omega_{i}<\omega<\omega_{f}$},\\
\text{$0$}, &\quad\text{for $\omega_{f}<\omega<\omega_{c}$},\\
\text{$+\frac{\omega}{2} \sqrt{R_{-}}$}, &\quad\text{for $\omega_{c}<\omega<\omega_{r}$},\\
\text{$0$}, &\quad\text{for $\omega>\omega_{r}$}.
\end{cases} \label{dicro_TL2.4}
\end{equation}

\begin{figure}[h]
	\centering
	\includegraphics[scale=0.65]{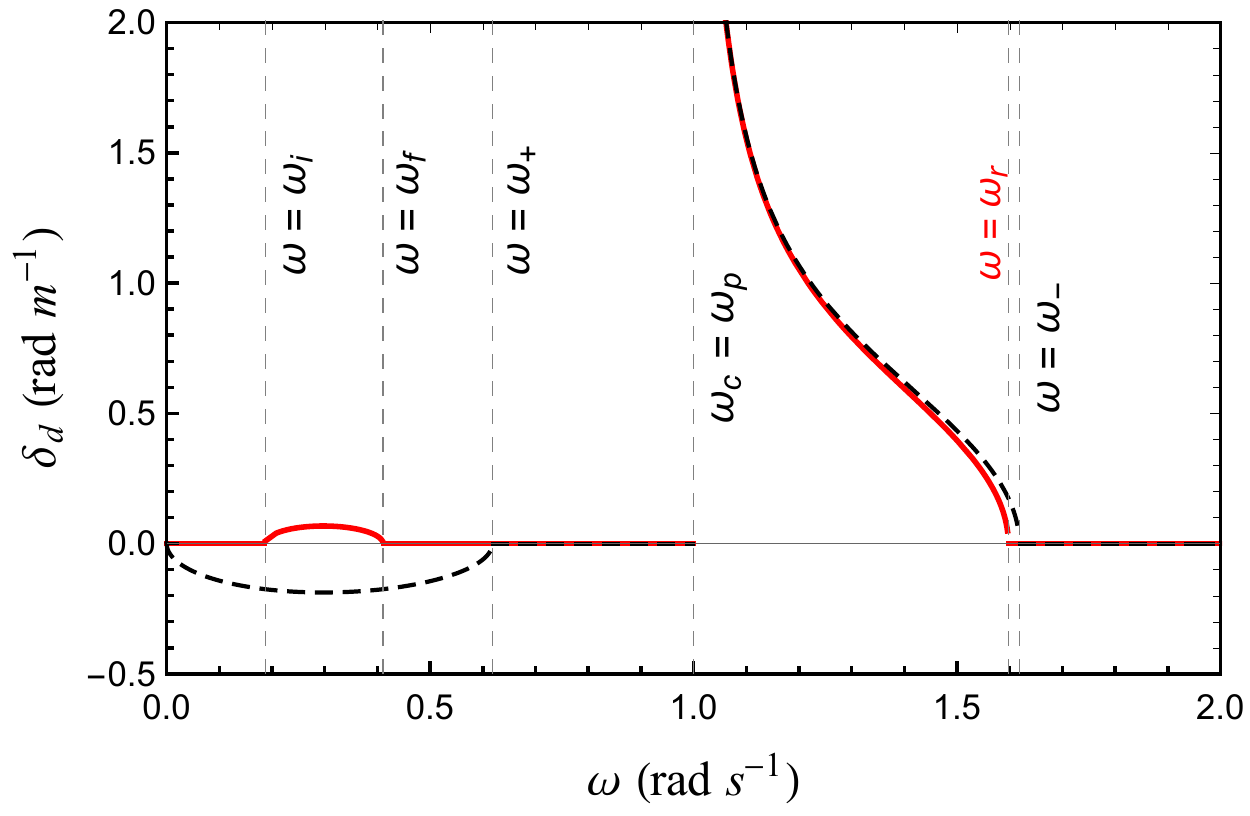}
	\caption{Plot of the dichroism coefficients (\ref{dicro_TL2.4}) associated to the refractive indices $n_{E}$ and $n_{R}$ [for the condition (\ref{cond2})]. The dashed line represents the usual dichroism coefficient (\ref{dicrousualcase}). Here, we have used $\omega_{c}=\omega_{p}$, $V_{0}=0.7\omega_{p}$, {and $\omega_{c}=1$~$\mathrm{rad}$~$s^{-1}$.}}
	\label{dicro_tlplot}
	
\end{figure}

The general behavior of the dichroism coefficient (\ref{dicro_TL2.4}) is illustrated in Fig.~\ref{dicro_tlplot}.

\section{Final remarks \label{conclusion}}

	In this work, we have examined the propagation of electromagnetic waves in a cold magnetized plasma in the context of the chiral MCFJ electrodynamics, describing the implied optical effects as well. We have adopted a MCFJ timelike background vector in order to represent the chirality factor that breaks the parity. Starting from the modified Maxwell equations and employing the usual methods, we obtained four modified refractive indices given by Eqs. (\ref{n-R-M-indices-1}) and (\ref{n-L-E-indices-1}), associated with circularly polarized propagating modes. Such indices were analyzed in detail in the Secs. \ref{secNR}-\ref{secNM}, where some of them exhibited significant modifications, as the index $n_{R}$, see Fig.~\ref{nRfig}. It presents a negative refraction behavior in the range $\omega_{c}<\omega<\omega_{-}$, in which it occurs propagation with absorption for $\omega_{c}<\omega<\omega_{r}$ and free (metamaterial) propagation for $\omega_{r}<\omega<\omega_{-}$. The usual counterpart index presents only pure absorption in this range. The {low-frequency} limit was investigated, there appearing propagating RCP and LCP helicons due to the presence of the chiral factor, $V_{0}$. Wave propagation orthogonally to the magnetic field was also investigated in Sec. \ref{sec5}, providing refractive indices and propagating modes modified by $V_{0}$.
	
Optical effects of this system, involving birefringence and dichroism, were discussed in Sec.~\ref{birefringence}, considering the refractive index $n_{L}$ and $n_{R}$ and $n_{E}$. In Sec.\ref{secRP}, the RP $\delta_{LR}$ was introduced, see \eqref{rptl}, exhibiting sign reversion at $\omega=\hat{\omega}$, for the conditions (\ref{cond}) and (\ref{cond2}). The RP $\delta_{ER}$ also exhibits  sign change at $\omega=\omega''>\omega_{c}$ for the condition (\ref{cond}), and  $\omega=\omega''<\omega_{c}$ under the condition (\ref{cond2}), as shown in Figs.~\ref{rptl1NE} and \ref{rptl1NE2}, respectively. The RP also increases with the frequency for $\omega>\omega_{c}$.  The reported RP reversal is not usual in cold plasmas, being reported in graphene systems \cite{Poumirol}, rotating plasmas \cite{Gueroult}, Weyl metals and semimetals with low electron density with chiral conductivity \cite{Pesin,Dey-Nandy}, and bi-isotropic dielectrics with magnetic chiral conductivity \cite{PedroPRB}. Comparing our results with the rotating plasma scenario of Ref.~\cite{Gueroult}, there appear differences. In the rotating plasma, the RP undergoes reversal and decays as $1/\omega^{2}$ for high frequencies. In the present case, the rotatory power tends to the asymptotical value $-V_{0}$, see \eqref{faradayTIMELIKE1}, or increases with $\omega$ when it involves the negative refraction index, see \eqref{faradayTIMELIKE2}. These distinct RP properties may provide a channel to optically characterize chiral cold plasmas, {being of experimental interest.}

Besides the nonconventional effect of reversion, the RP can also be enhanced when it is defined in the negative refraction zone. Such an enhancement occurs for $\delta_{ER}$, given in \eqref{rpeq2}, for $\omega>\omega_{+}$ (zone in which $n_{E}$ is negative), being a topic of interest in metamaterial plasmas \cite{Guo,Gao,Sakai,Sakai2}. Dichroism was examined in Sec.\ref{secDC}, where the coefficients $\delta_{dLR}$ and $\delta_{dER}$ have been shown to be non-null only in the range $\omega_{c}<\omega<\omega_{r}$, for the condition (\ref{cond})---see Figs.~\ref{dicro_tlplot1}, and in the intervals $\omega_{c}<\omega<\omega_{r}$, $\omega_{i}<\omega<\omega_{f}$, for the condition (\ref{cond2}), in accordance with Figs.~\ref{dicrotlplot2} and \ref{dicro_tlplot}.

{For a {cold-axion} dark matter, $\mathbf{\nabla }\theta={\bf{0}}$,  it holds $\mathbf{\nabla }\times \mathbf{B}-\partial _{t}\mathbf{E}=\mathbf{j}+V _{0} \mathbf{B}$,
	with $V_{0} =\partial_{t} \theta$, related to time dependence of the axion field.  If one considers $V_{0}$ constant, one has an effective chiral electrodynamics in the background of an axion field, as discussed in Sec. XI of Ref. \cite{Sikivie}. This is the scenario addressed in the present manuscript. On the other hand, a plasma interacting with a vibrating cold axion dark matter with frequency $\omega_{a}$ is an interesting problem that may open nice connections between cold plasmas and axion/dark matter systems, including the possibility of experiments involving dielectric haloscopes \cite{Millar} and tunable plasmas haloscopes \cite{Lawson, Millar-2}}.

	\begin{acknowledgments}
	
The authors express their gratitude to FAPEMA, CNPq, and CAPES (Brazilian research agencies) for their invaluable financial support. M.M.F. is supported by FAPEMA Universal/01187/18, CNPq/Produtividade 311220/2019-3 and CNPq/Universal/422527/2021-1. P.D.S.S is supported by  FAPEMA BPD-12562/22. Furthermore, we are indebted to CAPES/Finance Code 001 and FAPEMA/POS- GRAD-02575/21.

	\end{acknowledgments}

\end{document}